\documentclass[aps,pra,reprint,twocolum,amsmath,superscriptaddress,amssymb,showpacs,nofootinbib]{revtex4-1}
\usepackage{amsmath,braket,amssymb}
\usepackage[final]{graphicx}
\usepackage{subfigure}
\usepackage{xcolor}
\usepackage[english]{babel}
\usepackage[utf8]{inputenc}
\usepackage{color}
\usepackage{mathtools}
\usepackage{empheq}
\usepackage{tensor}
\usepackage{cancel}
\usepackage{amsmath}

\usepackage[colorlinks,citecolor=blue,linkcolor=magenta,
urlcolor=blue,hyperindex]{hyperref}

\usepackage[normalem]{ulem}
\normalfont

\definecolor{forestgreen}{rgb}{0.08, 0.4, 0.13}
\definecolor{darkBlue}{rgb}{0.08, 0.13, 0.4}

\definecolor{THc}{rgb}{0.9,0.3,0.2}

\begin{document}


\title{Triviality of quantum trajectories close to a directed percolation transition}

\author{Lorenzo Piroli}
\affiliation{Philippe Meyer Institute, Physics Department, \'{E}cole Normale Sup\'{e}rieure (ENS), Universit\'{e} PSL, 24 rue Lhomond, F-75231 Paris, France}

\author{Yaodong Li}
\affiliation{Department of Physics, Stanford University, Stanford, CA 94305, USA}

\author{Romain Vasseur}
\affiliation{Department of Physics, University of Massachusetts, Amherst, MA 01003, USA}

\author{Adam Nahum}
\affiliation{Laboratoire de Physique de l’\'{E}cole Normale Sup\'{e}rieure,
ENS, Universit\'e PSL, CNRS, Sorbonne Universit\'e,
Universit\'e Paris-Diderot, Sorbonne Paris Cit\'e, Paris, France}

\date{\today}

\begin{abstract}
We study quantum circuits consisting of unitary gates, projective measurements, and control operations that steer the system towards a pure absorbing state. Two types of phase transition occur as the rate of these control operations is increased: a measurement-induced entanglement transition, 
and a directed percolation transition into the absorbing state (taken here to be a product state).
In this work we show analytically that these transitions are generically distinct, 
with the quantum trajectories becoming disentangled before the absorbing state transition is reached, and we analyze their critical properties.
We introduce a simple class of models where the measurements in each quantum trajectory define an Effective Tensor Network (ETN)--- a subgraph of the initial spacetime graph where nontrivial time evolution takes place. 
By analyzing the entanglement properties of the ETN, 
we show that the entanglement and absorbing-state transitions coincide only in the limit of infinite local Hilbert-space dimension. 
Focusing on a Clifford model which allows numerical simulations for large system sizes, we verify our predictions and study the finite-size crossover between the two transitions at large local Hilbert space dimension.
We give evidence that the entanglement transition is governed by the same fixed point as in hybrid circuits without feedback.
\end{abstract}

\maketitle



\section{Introduction}

The competition between local interactions and local measurements
in a many-body system can give rise to a measurement-induced phase transition (MIPT)  \cite{skinner2019measurement, li2018quantum}. In the simplest setting, unitary quantum circuit dynamics is interspersed with local projective measurements, yielding a hybrid dynamics which is non-deterministic: different histories of the random measurement outcomes define distinct quantum trajectories. 
As the measurement rate is increased, the system undergoes a transition from a phase where typical quantum trajectories are volume-law entangled, to one in which they are area-law entangled.
This MIPT can also be understood as a dynamical purification transition, if the system is initialized in a maximally mixed state~\cite{gullans2020dynamical,bao2020theory,choi2020quantum}.

MIPTs can show various universality classes depending on the structure of the dynamics~\cite{potter2022entanglement,fisher2022random}. 
An important general question is what additional kinds of behavior are possible when the dynamics involves 
feedback, i.e. control operations~\cite{sierant2022dissipative,buchhold2022revealing,iadecola2022dynamical,friedman2022locality,friedman2022measurement}.
In the simplest setting, these adaptive operations are entirely local: each measurement is followed by a local unitary that is conditioned on that measurement outcome. 
Perhaps the simplest phase transition that can be engineered this way is a 
transition into an ``inactive''  product state, say $\ket{00\ldots 00}$ in the computational basis (for more complex steering protocols with nontrivial absorbing states, see e.g. Refs.~\cite{briegel2001persistent,
bolt2016foliated,
roy2020measurement,
piroli2021quantum,
tantivasadakarn2022hierarchy, mcginley2022absolutelystable} and references therein).
If the unitary part of the dynamics preserves the  $\ket{00\ldots 00}$ state, and if the feedback operations reset qudits to ``0'', then this is an absorbing state. As the rate of measurement and control operations is increased, the system can undergo an absorbing-state transition \cite{marcuzzi2016absorbing,lesanovsky2019non,carollo2022nonequilibrium,weinstein2022scrambling} in the directed percolation universality class \cite{lubeck2004universal,janssen2005field,henkel2008non}.  See Refs.~\cite{gillman2022using,popkov2020dissipative,gillman2022asynchronism} for examples of quantum models with absorbing-state transitions, and Refs.~\cite{tauber2014critical,hinrichsen2000non,henkel2008non} for reviews of classical nonequilibrium phase transitions.

Recently, the question  has been raised \cite{buchhold2022revealing} (see also~\cite{iadecola2022dynamical}) of the relation between 
the absorbing-state transition and the MIPT, with the hope of getting around the so-called post-selection bottleneck for experimental detection of MIPTs~\cite{noel2022measurement,koh2022experimental,fisher2022random}. 
This problem is also interesting \emph{per se}, allowing us to deepen our understanding of universal behavior in monitored dynamics, and could be relevant for the implementation of absorbing-state transitions in quantum devices, as recently reported in Ref.~\cite{chertkov2022characterizing}.

This work clarifies the interplay between the absorbing-state transition and the MIPT, and analyzes the corresponding critical properties (see also Ref.~\cite{2022arXiv221105162R} for recent results in that direction). 
We argue, using general 
properties of the statistical mechanical descriptions
of the two transitions, that they are, in general, distinct and unrelated to each other.
We design models where the two transitions can be fit into the same phase diagram with a single tuning parameter, and show both theoretically and numerically that they generically occur at different locations and are in different universality classes.

Our argument is based on two properties, which we  show microscopically for our models, and which we argue to generalize to other models after coarse-graining. 
First, the adaptive measurements in each quantum trajectory define an \textit{effective tensor network} (ETN) in which the time evolution takes place. 
This ETN is a subgraph of the initial spacetime graph, obtained by deleting bonds that are determined (by the measurements) to be in the ``trivial'' inactive state.
The ETN gives a simpler picture for the entanglement dynamics than the original quantum circuit, because of the second property:
The absorbing-state transition is also a directed percolation (DP) transition~\cite{henkel2008non} for the geometrical connectivity of the ETN. 
The fact that the ETN becomes only tenuously connected close to the percolation transition allows  us to show that the 
measurement transition occurs before the percolation transition.
This is done by considering the minimal cut properties of the percolation configurations and relating them to effective statistical mechanics models describing entanglement \cite{skinner2019measurement, RTNReplica, zhou2019emergent}.

In order to substantiate our predictions, we introduce a simple Clifford version of the model~\cite{gottesman1996class,gottesman1998heisenberg,aaronson2004improved}, defining a way of  ``flagging'' inactive qudits in order to define the ETN.
This allows us to obtain numerical results for large system sizes and simulation times which support the claim that the entanglement transition
is separated from the 
absorbing-state transition.
Instead, it occurs within the percolating phase, where the ETN has 2D connectivity
(as opposed to the fractal connectivity associated with the percolation critical point).
The limit of infinite on-site Hilbert-space dimension is an exception, so that for large finite Hilbert-space dimension the two transitions are close.
Going further, based on numerical results, we show that, while the absorbing-state transition is governed by the directed percolation fixed point, the entanglement MIPT appears to be in the same universality class as in the model without feedback, at least for Clifford.

We begin in Sec.~\ref{sec:hybrid_standard} by recalling a few elementary facts about entanglement and purification transitions without feedback. We then introduce the notion of adaptive and resetting measurements and absorbing states for the averaged dynamics in Sec.~\ref{sec:resetting}. In Sec.~\ref{sec:the_model} we introduce the simplified models studied in this work, and in Sec.~\ref{sec:arguments} we analyze  the relation between the absorbing-state and purification transitions theoretically. In Sec.~\ref{sec:Clifford_model_results} we report our numerical study of the Clifford model. In Sec.~\ref{sec:conclusions} we sketch why the main statements are more general than the specific models studied here, and present our conclusions. 

\section{Adaptive hybrid dynamics}

\begin{figure*}[t]
	\includegraphics[scale=0.6]{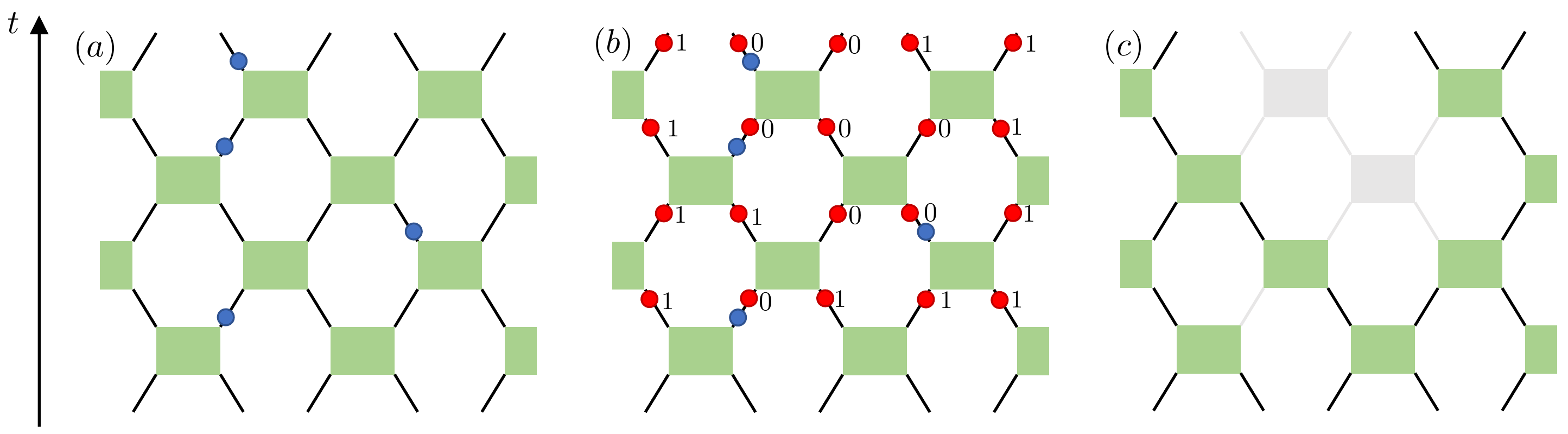}
	\caption{(a): Hybrid circuit composed of local unitaries and measurements.  The rectangles in the first two panels represent random unitary gates, arranged in a brickwork pattern. Blues dots denote single-qudit projective measurements, which are performed with some rate $p$. (b): Adaptive circuit defined in Sec.~\ref{sec:haar_random_model}. Blues dots represent the single-qudit resetting measurements~\eqref{eq:resetting_kraus}, while red dots represent measurements of the operator $\mathcal{P}_j$~\eqref{eq:p_operator}, which are performed at each time step. In a given quantum trajectory, each $\mathcal{P}_j$ has a definite value $0$ (``inactive'') or $1$ (``active''). $(c)$: By discarding the inactive bonds, we are left with a TN formed out of the active bonds, which are each now of dimension $q$, instead of ${q+1}$.}
	\label{fig:sketch}
\end{figure*}

\subsection{Hybrid dynamics and purification transitions}
\label{sec:hybrid_standard}

We consider a one-dimensional hybrid circuit acting on a set of $L$ qudits. We denote the local Hilbert space by $\mathcal{H}_j$, with $j=1,2,\ldots, L$ and ${{\rm{dim}}(\mathcal{H}_j)=q+1}$.
We will think of one of these states, $\ket{0}$, as the inactive state.
The circuit is composed of nearest-neighbor unitary gates $U_{j,j+1}$ and interspersed with local single-qudit measurement  processes, cf. Fig.~\ref{fig:sketch}(a). In general, these measurement processes are described by a set $\mathcal{M}^{(j)} = \{M^{(j)}_\alpha\}$ of Kraus operators satisfying
\begin{equation}
	\sum_{\alpha}[M_\alpha^{(j)}]^\dagger M^{(j)}_\alpha=\openone\,,
\end{equation}
where $\alpha$ and $j$ label the different outcomes and the qudit being considered, respectively. After the measurement, the density matrix of the system $\rho$ is updated as (dropping the label $j$)
\begin{equation}
	\rho\mapsto \frac{M_\alpha\rho M_\alpha^\dagger}{{\rm tr}[M_\alpha\rho M_\alpha^\dagger]}\,.
\end{equation}
Different measurement histories along the hybrid dynamics define the ensemble of quantum trajectories.

In the original setting introduced in Refs.~\cite{skinner2019measurement,li2018quantum,unitary2019chan}, measurements are performed with some finite probability $p$ at each time step, and are projective, {i.e.} 
\begin{equation}\label{eq:standard_measurement}
	M_{\alpha}=\ket{\alpha}\bra{\alpha}\,,
\end{equation}
where we denoted by $\ket{\alpha}$ the basis states for the local Hilbert space $\mathcal{H}_j$, with $\alpha=0,\ldots, q$. As the measurement rate $p$ is increased, the dynamics undergoes an entanglement MIPT or, equivalently, a dynamical purification transition~\cite{gullans2020dynamical,bao2020theory,choi2020quantum}. In the following, we will take this purification perspective, allowing for a slightly simplified discussion. In this protocol the system is initialized in a maximally mixed state
\begin{equation}\label{eq:initial}
	\rho(0)=\frac{\openone}{(q+1)^L}\,,
\end{equation} 
and we track the dynamics of the mixed state entropy
\begin{equation}\label{eq:entropy}
	S(t)=-{\rm Tr}\{\rho(t)\log[\rho(t)]\}\,.
\end{equation}
Because of the measurements, the entropy  ${S(t)}$ is, on average, a decreasing function of time $t$, and in the limit $t\rightarrow\infty$ (at fixed $L$) the state is purified. 
However,
the timescale for this extraction of entropy changes radically at the measurement transition  $p^{\rm ent}_c$ \cite{gullans2020dynamical,bao2020theory,choi2020quantum}.
For  $p<p^{\rm ent}_c$, the system retains an extensive entropy for a timescale that is exponentially long in $L$.
On times that are large compared to microscopic times, but short compared to this exponentially long timescale, the state  retains a finite ``steady state'' entropy density, $s(p)$, 
so that ${L^{-1} S(t)\simeq s(p)}$
\cite{gullans2020dynamical,bao2020theory,li2021statistical2,nahum2021measurement}.
On the other hand for $p>p^{\rm ent}_c$ 
the entropy density decays exponentially to zero on an order one timescale, and the steady state entropy density $s(p)$ vanishes.

\subsection{Resetting measurements and absorbing states}
\label{sec:resetting}

We wish to modify the measurement processes to allow for local feedback operations. 
This adaptive dynamics may have interesting new features compared with the above case~\cite{sierant2022dissipative,buchhold2022revealing,iadecola2022dynamical,friedman2022locality,friedman2022measurement}. Motivated by recent work~\cite{buchhold2022revealing}, we focus on \emph{resetting} operations that steer the system towards a pure fixed point.
For now we will consider the simplest possible example:
in the following section it will be convenient to slightly modify the model.
For this example we take the Kraus operators to be
\begin{equation}
	\widetilde{M}_{\alpha}=\ket{0}\bra{\alpha}\,.
	\label{eq:resetting_kraus}
\end{equation}
Physically this represents  a simple feedback operation, where we first perform a projective measurement in the computational basis, and then apply a local unitary operation mapping $\ket{\alpha}$ into $\ket{0}$, if the outcome is $\alpha\neq 0$. Eq.~\eqref{eq:resetting_kraus} describes this two-step process.

The resetting operation~\eqref{eq:resetting_kraus} steers the system towards the trivial state
\begin{equation}\label{eq:Omegadefn}
	\ket{\Omega}=\ket{0}_1\otimes \cdots\otimes \ket{0}_L\,.
\end{equation}
In order for this to be a fixed point of the hybrid dynamics, we need to restrict to unitary gates $U_{j,j+1}$ that act as the identity on the state $|00\rangle :=\ket{0}_j \otimes \ket{0}_{j+1}$, though they can act non-trivially on the subspace of $\mathcal{H}_j\otimes \mathcal{H}_{j+1}$ which is orthogonal to $\ket{00}$. Therefore, in the two-qudit basis 
\begin{equation}\label{eq:twositebasis}
\mathcal{B}=\{\ket{00}, \ldots, \ket{0q}, \ket{10},\ldots \ket{1q}, \ldots \}\,,
\end{equation}
we choose gates with block-diagonal form
\begin{equation}\label{eq:block_unitary}
	U=
	\begin{pmatrix}
		1 & 0\\
		
		0 & W
	\end{pmatrix}\,,
\end{equation}
where $W\in U(q(q+2))$ is a $q(q+2)\times q(q+2)$ unitary matrix. For now we assume that $W$ is drawn from the Haar random distribution over $U(q(q+2))$, independently for each pair of qudits and time step. 

The effect of the resetting measurement~\eqref{eq:resetting_kraus}, can be easily appreciated by looking at the dynamics of averaged observables 
\begin{equation}
\mathbb{E}\left\{{\rm Tr} [\rho(t)\mathcal{O}_{A}  ]\right\}={\rm Tr} \{ \mathbb{E}[\rho(t)]\mathcal{O}_{A} \}\,.
\end{equation}
Here $\mathbb{E}[\cdot]$ denotes the ensemble average over all the measurement outcomes and unitary gates, while $\mathcal{O}_A$ is some observable supported over the region $A$. As discussed in detail {e.g.} in Ref.~\cite{fisher2022random,potter2022entanglement,jian2020measurement}, the averaged density matrix $\mathbb{E}[\rho(t)]$ undergoes a quantum-channel dynamics, where one- and two-qudit quantum channels~\cite{nielsen2002quantum} are applied in sequence according to a brickwork structure. 

For contrast, first consider the non-adaptive dynamics in Sec.~\ref{sec:hybrid_standard}.
A simple projective measurement~\eqref{eq:standard_measurement} corresponds to a completely dephasing channel acting on qudit $j$
\begin{equation}\label{eq:dephasing}
	\rho\mapsto \mathcal{E}^{(j)}_D[\rho]=\sum_{\alpha} \tensor[_j]{\bra{\alpha}}{}\rho  \tensor[]{\ket{\alpha}}{_j}\otimes  \ket{\alpha}\bra{\alpha}_j\,.
\end{equation}
The effect, after averaging, 
of a completely generic Haar unitary $U_{j,j+1}\in U((q+1)^2)$
is, in a more informal language, to eliminate coherences
(so that after the first layer of gates, the averaged density matrix reduces to a ``classical'' probability vector)
and also to equalize the occupation probabilities of all of the $(q+1)^2$ basis states for the pair of sites (\ref{eq:twositebasis}).
In the absence of measurements,
such generic gates drive the averaged density matrix to the infinite-temperature state. Since $\mathcal{E}^{(j)}_D(\openone)=\openone$, the dephasing channel arising from measurement does not compete with the unitary dynamics, and the averaged  quantum-channel evolution remains trivial for all values of the measurement rate $p$.

On the other hand, if we  choose the resetting measurements~\eqref{eq:resetting_kraus} and restrict to gates of the form~\eqref{eq:block_unitary}, the averaged channel dynamics maps to a nontrivial classical stochastic process.
The restricted unitary gates (\ref{eq:block_unitary})
still eliminate coherences, meaning that  the averaged density matrix is diagonal and again reduces to a classical probability vector for the basis states.
The application of a unitary amounts after averaging to a Markovian update of this probability vector.
However, this update now  redistributes probability only among the ${(q+1)^2-1}$ states that are orthogonal to $\ket{00}$.
The measurement of qudit $j$ corresponds to a channel that deactivates that qudit,
\begin{equation}\label{eq:resetting_channel}
	\rho\mapsto \mathcal{E}^{(j)}_R[\rho]=\left(\sum_{\alpha} \tensor[_j]{\bra{\alpha}}{}\rho  \tensor[]{\ket{\alpha}}{_j}\right)\otimes  \ket{0}\bra{0}_j\,.
\end{equation}
This channel is not unital, {i.e.} $\mathcal{E}_R[\openone]\not\propto\openone$. In fact, we see that $\ket{\Omega}$ in Eq.~\eqref{eq:Omegadefn} is the only state left invariant by $\mathcal{E}^{(j)}_R$ for all $j$. In addition, $\ket{\Omega}$ is also fixed by the unitary part of the dynamics, due to the block structure of~\eqref{eq:block_unitary}, making it an \emph{absorbing state} for the averaged evolution. 
However, the resetting measurements compete with the unitary dynamics which, starting from a generic initial state, would produce an equilibrium state with 
a positive density of active qudits.
Accordingly, the  averaged dynamics shows  a transition for a critical value of the measurement rate $p^{\rm abs}_c$, which is detected by the order parameter
\begin{equation}\label{eq:order_parameter}
	n(t)=\frac{1}{L}\sum_{j=1}^L{\rm Tr}\left\{ \mathbb{E}[\rho(t)] \mathcal{P}_j\right\}\,,
\end{equation}
where 
\begin{equation}\label{eq:p_operator}
	\mathcal{P}_j=\openone-\ket{0}\bra{0}_j\,
\end{equation}
is the ``occupation number'' of active qudits.
For ${p<p^{\rm abs}_c}$ the stable steady state has
\begin{equation}
n_p \equiv \lim_{t\rightarrow\infty}\lim_{L\rightarrow \infty} n(t)  >0\,,
\end{equation}
while for ${p \geq p^{\rm abs}_c}$ this steady-state density vanishes. 

Importantly, the quantum-channel dynamics does not provide full information on the  entropy of quantum trajectories,  because $S$ in~\eqref{eq:entropy} is a non-linear function of $\rho$. The steady-state entropy density $s(p)$ must of course vanish in the inactive phase (since individual trajectories, like the averaged state, are reset to the trivial inactive state),
but we expect it to be nonzero for small $p$, implying an entanglement transition at some critical value $p^{\rm ent}_c$, 
with $p^{\rm ent}_c\leq p^{\rm abs}_c$. 

Below, we clarify the relation between these transitions for a class of circuits that allows us both to develop robust analytical arguments and to provide numerical results for large system sizes and simulation times.
We will also argue in Sec.~\ref{sec:conclusions} that our conclusions extend to more general models with an absorbing state transition into a pure product state (for example models of the type described above). 
The concrete models we consider below have
an additional simplification with respect to the circuit presented in this section and to more general types of resetting dynamics. 
This simplification is that the quantum channel transition can be related
very directly to the connectivity of the ETN associated with a typical quantum trajectory. In the following, we will consider both Haar-random and random Clifford circuits.

\section{The models}\label{sec:the_model}

In this section, we introduce two types of models, proceeding in two steps. First  we   define a model in which the ``occupation number''  $\mathcal{P}_j$
is measured for every qudit in every time step.   
This is the setting where the definition of the ETN is the most straightforward.
However, this model does not have an obvious (efficiently simulable) Clifford generalization.
Therefore we next show how to define the ETN for a slightly broader class of models, in which the experimentalist does \textit{not} measure the occupation numbers of all of the qudits.

\subsection{The Haar-random circuit}\label{sec:haar_random_model}

The first model we consider is a simplification of the circuit introduced in Sec.~\ref{sec:resetting} and is represented pictorially in Fig.~\ref{fig:sketch}(b). It is made up of the following three ingredients:
\begin{enumerate}
	\item We apply a brickwork pattern of  two-site unitary gates $U_{j,j+1}$ with the block structure~\eqref{eq:block_unitary}. They act on qudits with $q+1$ states, and $W$ in~\eqref{eq:block_unitary} is Haar-random;

 	\item After each layer of gates, independently for each site, we perform a resetting operation~\eqref{eq:resetting_kraus} with probability $p$;
    
    \item Next, we measure the occupation number  $\mathcal{P}_j$ defined in~\eqref{eq:p_operator} for every qudit.
 
\end{enumerate}
Since $\mathcal{P}_j$ is measured for all qudits $j$ at each  time step, the resulting entanglement dynamics is only non-trivial for $q\geq 2$. 
In a given quantum trajectory, $\mathcal{P}_j$ has a definite value, either $0$ or $1$,
for every link of the  TN associated with the circuit, cf. Fig.~\ref{fig:sketch}(b). 
This integer plays the role of a classical ``flag'', 
distinguishing active and inactive qudits. In a given quantum trajectory, every link of the circuit that is  flagged as inactive has a projection operator $\ket{0}\bra{0}$ on it. 
These can essentially be deleted from the tensor network: 
the state at time $t$, 
in a given trajectory, 
is determined by a reduced tensor network in which various bonds and unitaries are eliminated. 
This is illustrated in Fig.~\ref{fig:sketch}(c).
We defer a more detailed discussion to Sec.~\ref{sec:arguments}.

We note that the measurements of all the  $\mathcal{P}_j$ are not needed in order to observe an absorbing-state or a purification transition. Next we discuss models without this step.

\subsection{``Cliffordizable'' model}
\label{sec:clifford_model}

Clifford circuits have played an important role in developing our understanding of entanglement transitions, see {e.g.}~\cite{PhysRevB.100.134306, turkeshi2020measurement,li2021conformal,zabalo2022operator}. 
Ensembles of random Cliffords form a $2$-design~\cite{gottesman1998heisenberg}, {i.e.} they agree with  the statistical properties of Haar-random unitaries up to the second moment, which means that some properties (such as the averaged quantum channel dynamics) coincide between Clifford and Haar models. On the other hand, they can be simulated efficiently using the stabilizer formalism~\cite{gottesman1996class,gottesman1998heisenberg,aaronson2004improved}. 

Therefore it will be convenient to define a  hybrid Clifford circuit displaying both an absorbing-state and a purification transition. 
The model introduced in Sec.~\ref{sec:haar_random_model} relied on the possibility of writing down non-trivial unitary gates
with the block diagonal form in Eq.~\eqref{eq:block_unitary}.
As a consequence, this model is not straightforwardly ``Cliffordized'', because Clifford unitaries with the block structure~\eqref{eq:block_unitary} are trivial. 
In order to get around this problem, we introduce a different model with the same underlying simplifications. 
This is based on the concept of ``flagged qudits'',
and can be defined either with Haar-random or random Clifford gates: 
because of the 2-design property of each ensemble, the averaged quantum channel dynamics will be the same in either case, but the entanglement transition will differ. For concreteness, we restrict to the Clifford case. Importantly, this model does not involve a splitting of the Hilbert space $\mathcal{H}_j\otimes \mathcal{H}_{j+1}$, which could make it more suitable for future applications. 

The idea is to introduce a \emph{flag} variable for each qudit, 
which labels its status as ``inactive'' or ``active''.
This piece of information is encoded in a classical bit $f_j=0,1$. We can think of $f_j$ as expressing the experimentalist's (partial) knowledge of the state of the qudit at a given time.
In the  model of Sec.~\ref{sec:haar_random_model} above, $f_j$ could be set by the direct measurement of $\mathcal{P}_j$ at each time step. 
In the model below the experimentalist does not have quite as much information, 
but can still assign flags by a modified rule.

We again consider a discrete dynamics where qudits are updated in pairs in the usual brickwork pattern. At each  time step, and for every pair of qudits to be updated, we have the following operations:
\begin{enumerate}
	\item Unitary updates:
	\begin{enumerate}
	\item given the qudits $j$ and $j+1$, we apply a unitary gate if at least one of them is active ({i.e.} $f_{j}=1$ or $f_{j+1}=1$); the gate is chosen to be a random Clifford gate;
	\item if both of them are inactive ($f_j=f_{j+1}=0$), then no unitary is applied;
	\item if a unitary is applied, then both qudits are set to ``active'', $f_j=f_{j+1}=1$;
	\end{enumerate}
	\item Measurement process:
	\begin{enumerate}
		\item After each row of unitaries, each qudit $j$ undergoes a resetting measurement  (\ref{eq:resetting_kraus}) with probability $p$;
		\item if the qudit undergoes the resetting operation, we then set its flag to ``inactive'', $f_j=0$.
	\end{enumerate}
\end{enumerate}
We note that (if desired) the classical flag could be incorporated into the Hilbert-space structure, enlarging $\mathcal{H}_j\to\mathcal{H}_j\otimes \mathcal{F}_j$, where $\mathcal{F}_{j}$ is generated by $\ket{f_j}$ with $f_j=0,1$. The above rules then define a dynamics in which the state of the flags is always ``classical'', i.e. they are never in a superposition.
By definition, the initial state is taken to have $f_j=1$ for every qudit.

\section{Separating absorbing-state and purification transitions}
\label{sec:arguments}

We move on to analyze the models introduced in the previous section. We show that they display both an absorbing-state and a purification transition and that $p^{\rm ent}_c<p^{\rm abs}_c$, with $p^{\rm ent}_c=p^{\rm abs}_c$ only in the limit $q\to\infty$. We will initially focus on the Haar random circuit defined in Sec.~\ref{sec:haar_random_model}, and then describe at the end how the same arguments hold for the model introduced in Sec.~\ref{sec:clifford_model}, cf. also Sec.~\ref{sec:Clifford_model_results}.

\subsection{The directed percolation transition}

We first focus on the quantum-channel dynamics and study the evolution of the order parameter~\eqref{eq:order_parameter} or equivalently, of its local version
\begin{equation}\label{eq:n_j}
	n_j(t)={\rm Tr}\left\{\mathbb{E}[\rho(t)]\mathcal{P}_j\right\}\,,
\end{equation}
where the average is over all measurement locations/outcomes and over all unitary gates. 
For the model of Sec.~\ref{sec:haar_random_model}, and in a given quantum trajectory, $\mathcal{P}_j$ has a definite value, $0$ or $1$, after every round of measurements (we will abuse notation and denote by $\mathcal{P}_j$ both the operator and the numerical value after a measurement). Using standard techniques~\cite{oliveira_generic_2007,harrow_random_2009,nahum2018operator,vonKeyserlingk2018operator,khemani2018operator}, 
it is easy to see that the probability for given  values of
$\mathcal{P}_j$, after averaging over the unitaries, 
is given by a simple Markov process for $\mathcal{P}_j$, which is defined as follows. 
Consider two input qudits $(j,j+1)$, undergoing a unitary gate and subsequent measurement operations. If $\mathcal{P}_{j}=\mathcal{P}_{j+1}=0$ for the input qudits, then the output qudits are inactive with probability $1$. Conversely, suppose at least one of the input qudits is active. Then, denoting by $p[(a,b)]$ the probability that $(\mathcal{P}_j,\mathcal{P}_{j+1})=(a,b)$ after the measurements, we have
\begin{subequations}
\label{eq:outputprobs}
\begin{align}
p[(0,0)]& = \frac{p(2+pq)}{2+q}, \\
p[(0,1)]&= p[(1,0)] = \frac{(1-p)(1+pq)}{2+q},\\
p[(1,1)]& = \frac{q(1-p)^2}{2+q}\,.
 \end{align}
\end{subequations}
If we visualize the tensor network associated with the circuit as a rotated square lattice, where unitary gates represent the nodes,  Eqs.~\eqref{eq:outputprobs} define a classical Markov process in which the degrees of freedom are occupation numbers on the bonds. This model is a slight variation of the standard bond DP problem, which is well studied in the classical literature~\cite{henkel2008non}.
In this standard model, the two outputs of an active node are \textit{independently} chosen to be inactive with probability $\tilde p$,  defining a Markov process for the bonds with output probabilities analogous to \eqref{eq:outputprobs} of the form:
\begin{subequations}
\label{eq:standardoutputprobs}
\begin{align}
	p[(0,0)] &=\tilde p^2, \\
	p[(0,1)]&= p[(1,0)] =\tilde p (1-\tilde p),\\
	p[(1,1)] &= (1-\tilde p)^2\,.
\end{align}
\end{subequations}
For this problem, the critical value of $\tilde p$, $\tilde{p}_c$, is known to high accuracy~\cite{jensen1999low} and is equal to
\begin{equation}\label{eq:pc}
\tilde p_c= 0.355299814(5).
\end{equation}

It is easy to check that in the limit
$q\rightarrow\infty$ 
the probabilities in \eqref{eq:outputprobs} reduce to the form~\eqref{eq:standardoutputprobs}, with $\tilde p = p$. 
Therefore, in this limit the Markov process describing the dynamics of $n_j(t)$ coincides with standard bond DP, and the critical measurement rate is given by~\eqref{eq:pc}. For finite $q<\infty$ the two output bonds are correlated, but we still expect that  $n_j(t)$ displays a DP transition. The strength of this correlation in the outputs is of order $q^{-2}$. 
Namely,  if we neglect terms of order $q^{-2}$ then the probabilities in Eq.~\eqref{eq:outputprobs} can be written in the form Eq.~\eqref{eq:standardoutputprobs}, with $\tilde p = {p+(1-p)/q}$. Therefore, we may obtain an estimate of the critical point in model \eqref{eq:outputprobs} that is accurate up to order $q^{-1}$ by setting ${p+(1-p)/q}$ equal to $ \tilde p_c$, yielding for the model in Sec.~\ref{sec:haar_random_model}:
\begin{equation}
p^{\rm abs}_c  \simeq 0.3553 - \frac{0.6447}{q} + {O}(q^{-2}).
\end{equation}
We see that the effect of having a finite $q<\infty$ is to lower the value of the critical measurement rate $p^{\rm abs}_c$. We have verified that these conclusions are consistent with classical numerical simulation of the Markov process defined by~\eqref{eq:outputprobs}. In conclusion, we find that the quantum-channel dynamics displays an absorbing-state transition, in the universality class of DP.

The Clifford model of Sec.~\ref{sec:clifford_model} also undergoes a DP transition. 
In fact, in this model the dynamics of the flags reduces \emph{exactly} to the standard  bond DP model, as we discuss towards the end of the next section. As a result
\begin{equation}
p^{\rm abs}_c =\tilde p_c\,, \quad \forall q\,.
\end{equation}
where $\tilde p_c$ is defined in Eq.~\eqref{eq:pc}.

\subsection{The purification transition}
\label{sec:MIPT}

\begin{figure}
	\includegraphics[scale=0.5]{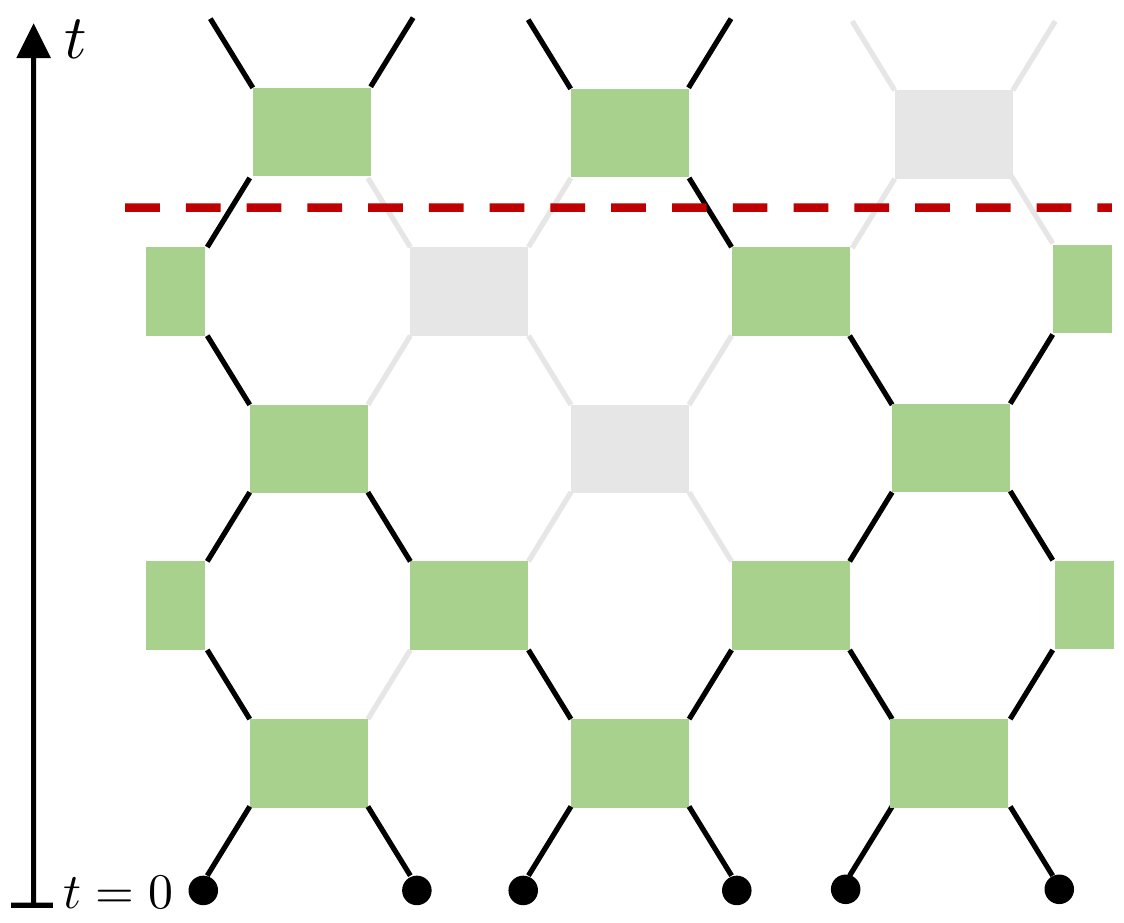}
	\caption{Schematic representation of the minimal-cut picture for purifying dynamics. Black dots correspond to the initial infinite temperature state $\rho_j(0)=\openone_j/(q+1)$. Black and gray links correspond to active and inactive bonds, respectively. Analogously, green and gray rectangles denote gates acting on active and inactive pairs of qudits. At time $t$, the entropy $S(t)$ is only determined by the active links and gates. In the limit of infinite onsite Hilbert space dimension $q\to \infty$, entanglement properties are dictated by the minimal-cut picture~\cite{hayden2016holographic,nahum2017quantum,skinner2019measurement}, namely the entropy  is proportional to the number of active links to be cut in order to separate the top and bottom boundaries of the ETN. The red dashed line corresponds to such a  minimal cut.}
	\label{fig:minimal_cut}
\end{figure}

\begin{figure*}
    \centering
    \includegraphics[width=.9\textwidth]{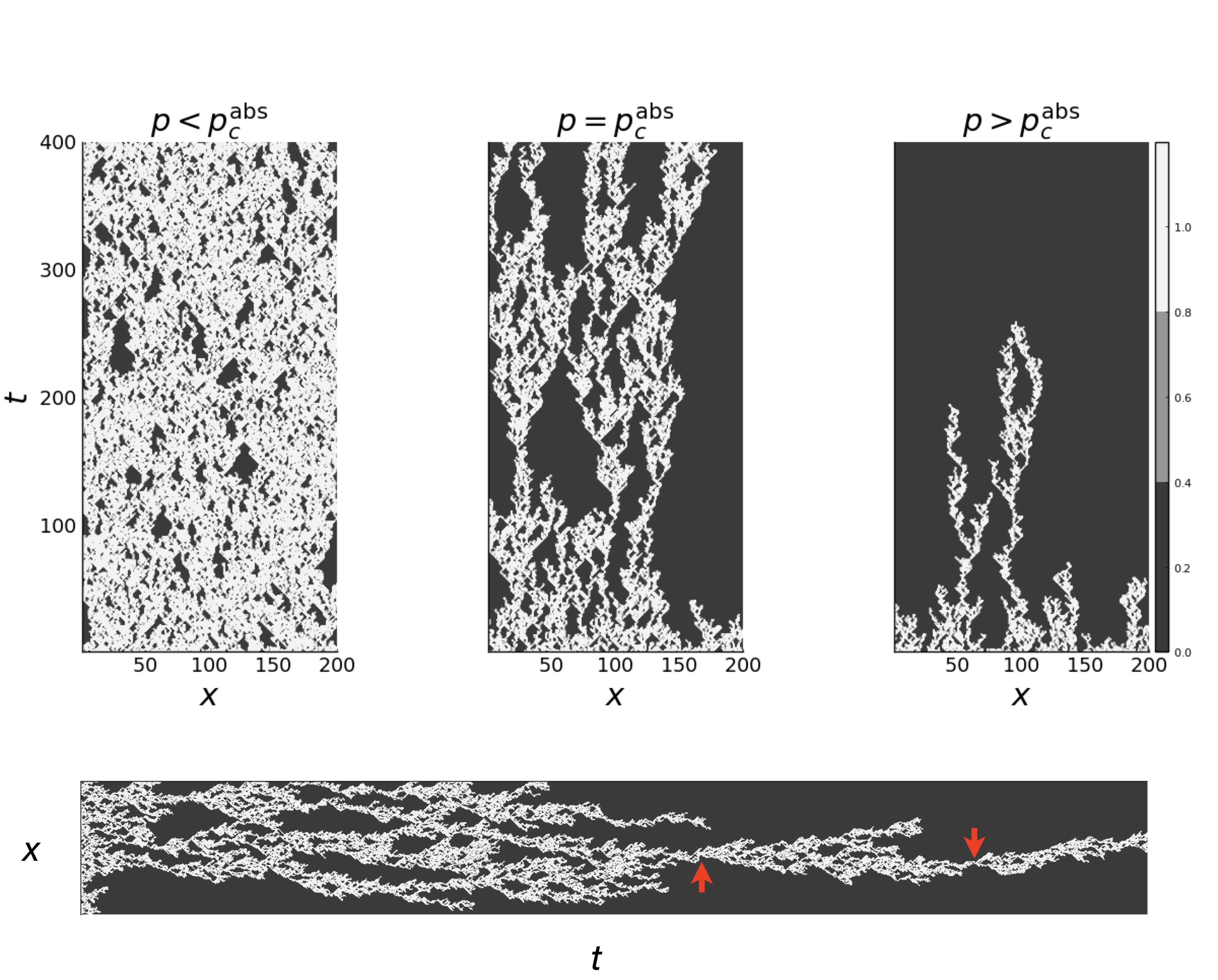}
    \caption{(Upper panel) Typical configurations of DP with $L = 200$ and $t \leq 400$, as produced by the \emph{standard} Markov process in Eq.~\eqref{eq:standardoutputprobs}.
    Here active sites are in white, and inactive sites are in black.
    (Lower panel)
    A configuration at the critical point, shown sideways, to access its shape at long times ($L = 200$, $t \leq 1600$).
    We highlight two red bonds with  red arrows.
    }
    \label{fig:dp-config}
\end{figure*}

We now argue that the Haar-random model also displays a purification transition for a value of $p^{\rm ent}_c<p^{\rm abs}_c$, the two coinciding only for $q\to\infty$ (these conclusions then extend to the Clifford model). 

First, it is important to note that the values of $\mathcal{P}_j$ not only determine the order parameter for the quantum-channel transition, but they also define the effective tensor network (TN) on which the dynamics take place. To be precise, let us fix the total simulation time $t$, and focus on the output state for a given history of the measurement outcomes, starting from the maximally mixed initial state~\eqref{eq:initial}. This output state is described by a TN, whose bonds are also labelled with $0$s and $1$s, depending on the values of $\mathcal{P}_j$. 
As shown in Fig.~\ref{fig:sketch}(c), we may effectively erase the inactive bonds and the nodes acting on pairs of them, because they correspond to trivial states and operations. The resulting active bonds and nodes form an ETN defining the non-trivial part of the output-state density matrix at time $t$. Note that this tensor network features several types of node tensors. For instance, a node with $4$ active bonds is a non-unitary tensor given by a $q^2$-dimensional block of the original random unitary. 

Crucially, the connectivity properties of this TN depend on $p$, because they are described by the bond DP problem. Consider the case $p<p^{\rm abs}_c$ for a large system and large time, with $t$ scaling polynomially in $L$.
We then have a cluster of active bonds connecting the initial state with the boundary at time $t$. The active cluster encloses elongated regions of inactive bonds, whose maximal space (time) dimension is of the order the spatial (temporal) correlation length $\xi_\perp$ ($\xi_\parallel$)~\cite{henkel2008non} (see Fig.~\ref{fig:dp-config}). 
These correlation lengths diverge as
$p^{\rm abs}_c$ is approached, with 
$\xi_\parallel\sim \xi_\perp^z$ and $z>1$ (see Sec.~\ref{sec:absorbingqinfinity}).
Conversely, if $p>p^{\rm abs}_c$, the bonds at time $t$ will all be inactive . 

We now wish to estimate the entropy of the output state at such large times $t$ (polynomially large in $L$) for ${p \lesssim p^{\rm abs}_c}$. 
As a warm-up, we first consider the so-called max-entropy $S_0(t)$.
As discussed in~\cite{nahum2017quantum,skinner2019measurement}, this quantity is particularly simple, as it can be obtained by an elementary ``minimal-cut'' picture: $S_0(t)$ is proportional to the minimal number of active links which have to be cut in order to separate the output qudits from the input ones, cf. Fig.~\ref{fig:minimal_cut}. Therefore, $S_0(t)$ is uniquely determined by the connectivity properties of the ETN, and governed by the physics of DP (or more precisely by the min-cut properties of DP configurations). In particular, $S_0(t)$ undergoes a transition from volume-law to area-law at the DP critical point $p^{\rm abs}_c$.
The max-entropy itself is not a very physically meaningful quantity (it is not stable to infinitesimal perturbations of the state). 
However,
in the limit $q\rightarrow\infty$ of infinite bond dimension,
the von Neumann and higher R\'enyi entropies are also given by the min-cut result
\cite{hayden2016holographic,nahum2017quantum}.
Therefore in the $q\rightarrow\infty$
limit the entanglement 
properties of the state are determined by the min-cut picture applied to the ETN, and the entanglement transition occurs at the DP critical point: as $q \to \infty$, we have $p^{\rm abs}_c=p_c^\text{ent}$.  

Next we wish to perturb around this limit of large $q$.
For finite $q$, 
the minimal-cut picture is not adequate to capture the behavior of $S(t)$,
but a closely related geometrical picture can be obtained at large scales, 
with domain-wall degrees of freedom taking the role of the min-cut.
For hybrid dynamics, this was shown in Refs.~\cite{bao2020theory,jian2020measurement}, extending the theory developed for unitary circuits in Refs.~\cite{nahum2017quantum,nahum2018operator,zhou2019emergent,zhou2020entanglement} and for random tensor networks in Refs.~\cite{hayden2016holographic,RTNReplica}. In these works, the dynamics was mapped onto an effective statistical mechanics model,  allowing one to relate the entropy to a domain wall free energy, receiving both ``energetic” and ``entropic” contributions. For our purposes we will not need the details of these constructions: 
to determine which phase we are in, it will be sufficient to make a  comparison of energy versus entropy
(analogous to one which may be made in the non-adaptive circuit at large $q$ \cite{skinner2019measurement})
that only relies only on two facts:
 ($1$) the degrees of freedom in the effective model are discrete, so that the domain walls are line-like objects; and ($2$) the energy cost per unit length of a domain wall is $\log q$ (in a convention where ``$k_B T$'' in the effective statistical mechanical model is~1).
 
To this end, we fix $\delta p=p^{\rm abs}_c-p$ small, so that we are just inside the percolating phase,
and focus on a sub-region of the TN with space and time dimensions $\ell\sim \xi_\perp$ and $\tau\sim \xi_\parallel$. We recall
\begin{equation}
	\xi_{\perp}\sim \delta p^{-\nu_\perp}, \quad 	\xi_{\parallel}\sim \delta p^{-\nu_\parallel}, 
\end{equation}
where $\nu_\perp$ and $\nu_\parallel$ are the critical exponents of DP, which are known numerically to high precision~\cite{henkel2008non}, cf. Eq.~\eqref{eq:exponents}. 
Since $\ell$ and $\tau$ are chosen to be of the same order of magnitude as the correlation lengths, the structure of the cluster of active links within this region is similar to that of a percolating cluster at the critical point. Such clusters have been studied in detail in the DP literature~\cite{frojdh2001directed}. An important result quantifies the number $N_{\rm red}(t)$ of the so-called \emph{red bonds}, {i.e.} the active links which, if removed, make all the bonds at time $t$ inactive~\cite{coniglio1981thermal,huber1995distribution, frojdh2001directed}. An illustration of the red bonds is given in Fig.~\ref{fig:dp-config}. Using a scaling argument, it was shown that, at criticality, and for a rectangular sample with the above geometry,
\begin{equation}
N_{\rm red}(\tau)\sim \tau^{1/\nu_\parallel}. 
\end{equation}
Therefore, the cluster inside the rectangular space-time region of dimensions $\ell$ and $\tau$ has order $\tau^{1/\nu_\parallel}\sim 1/|\delta p|$ bonds where it can be 
cut (horizontally) so as to disconnect the top and the bottom of the region.

We now consider the free energy cost of domain walls in the degrees of freedom that live on the ETN in the effective model. 
Consider a segment of such domain wall, 
oriented roughly horizontally, that has to cut through the spacetime patch under consideration. 
The minimal ``energetic cost'' for the domain wall is given by passing through one of the red bonds, and this gives a cost of $\log q$.
On the other hand, the domain wall 
has a choice of ${N_{\rm red}(\tau)\sim 1/|\delta p|}$ locations to cut the region,
giving a configurational entropy ${\sim \ln N_{\rm red}(\tau)}$.
This allows us to fix the leading terms in the domain wall free energy, coarse-grained on the scale of our spacetime patch:
\begin{equation}\label{eq:Festimate}
\mathcal{F} = \ln q- \ln (1/|\delta p|). 
\end{equation}
We see that if $q |\delta p|\gg 1$, the domain wall free energy is large and positive \cite{skinner2019measurement}. 
The effective model is then ordered, and the hybrid dynamics is in an entangling phase with volume-law entanglement \cite{bao2020theory,jian2020measurement}.
In the purification protocol, the steady-state entropy density $s(p)$ (Sec.~\ref{sec:hybrid_standard})
is given by entanglement ``membrane tension'' set by the free energy per unit length of the above domain wall,~${s(p)\sim \mathcal{F}/\xi_\perp}$.

Conversely, if $q |\delta p|\ll 1$,
so that the above free energy estimate is large and negative, we expect that domain walls proliferate and the effective statistical mechanics model is disordered, which corresponds to an area-law (or pure) phase. 
Put differently, if
\begin{equation}\label{eq:inequality}
|\delta p|	\ll 1/q\,,
\end{equation}
the system is in an area law phase, meaning that the purification transition takes place at some $p^{\rm ent}_c<p^{\rm abs}_c$. 
Going further, Eq.~\eqref{eq:Festimate} indicates that
\begin{equation}
p^{\rm abs}_c - p^{\rm ent}_c \sim \frac{1}{q}.
\end{equation}
In conclusion, we see that the absorbing-state and purification transitions are always separated, only coinciding when $q\to\infty$, as anticipated. The above discussion also indicates that $1/q$ is a relevant perturbation of the $q=\infty$ critical fixed point,  which will change the universality class of the entanglement transition.

The above conclusions can be extended to the models introduced in Sec.~\ref{sec:clifford_model}, as we now briefly discuss. 
First of all, we can again define an ETN, by eliminating all the bonds whose classical flag $f_j$ is equal to zero.
Now, however, the order parameter 
\eqref{eq:n_j}
is not equal to the 
density of nonzero flags:
while $f_j=0$ implies $\mathcal{P}_j=0$,
a bond with $f_j=1$ need not have a well-defined eigenvalue of the operator  $\mathcal{P}_j$. 
However, there is a simple relation between the two quantities.
Let us define the ``classical density''
\begin{equation}\label{eq:classical_density}
n^{\rm cl}(t)=\frac{1}{L}\sum_{j=1}^L \mathbb{E}[f_j(t)]\,.
\end{equation}
It is easy to show that the classical flags undergo a Markov process, defined on the links of TN associated with the circuit. In fact,
the transition probabilities are exactly 
those of the standard bond DP model:
given that a node is active
(i.e. that at least one of its input qudits is active), 
the output transition probabilities are given by~\eqref{eq:standardoutputprobs}, after replacing $\tilde{p}$ with $p$. Therefore, the flag dynamics undergoes a DP transition, whose critical measurement rate is $p^{\rm abs}_c=\tilde{p}_c$ for all values of $q$ [with $\tilde{p}_c$ defined in Eq.~\eqref{eq:pc}], and with order parameter given by $n^{\rm cl}(t)$. Crucially, it is easy to show
\begin{align}
n(t)=\frac{q}{q+1}n^{\rm cl}(t)\,,
\end{align}
where $n(t)$ is defined in~\eqref{eq:order_parameter}. 
Therefore, $n(t)$ is determined by the dynamics of the the classical flags, which undergo an absorbing-state transition at $p^{\rm abs}_c$.

In addition, the classical flags define the effective TN in which the dynamics take place, so we have once again that the absorbing-state transition corresponds to a DP transition for the geometrical connectivity of the ETN. In conclusion, we can directly repeat all the arguments presented for the Haar-random model.

In the next section, we will test our predictions via numerical computations. We will restrict to the Clifford model of Sec.~\ref{sec:clifford_model}, which allows us to reach large system sizes and simulation times. 

\section{Numerical results for the flagged Clifford circuit 
\label{sec:Clifford_model_results}}

\begin{figure}[b]
    \centering
    \includegraphics[width=.48\textwidth]{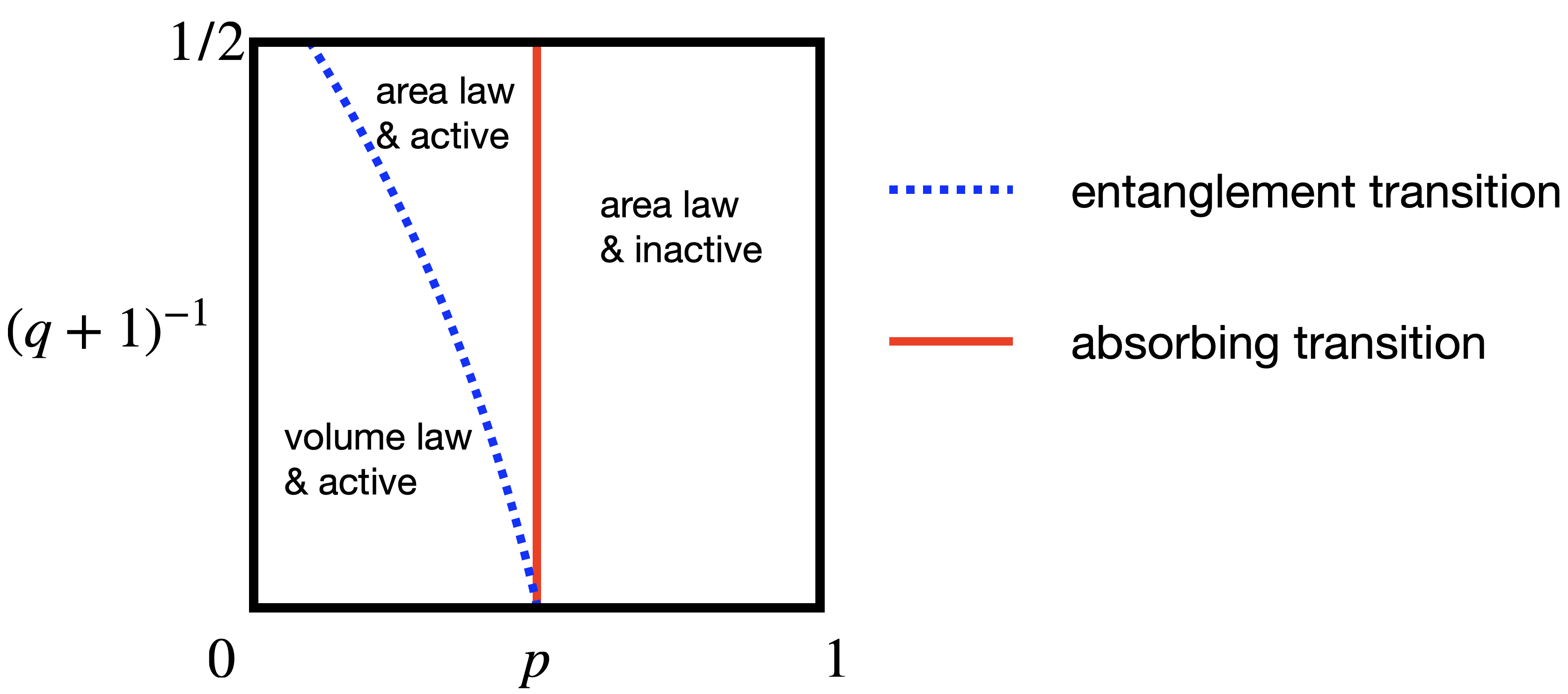}
    \caption{Schematic phase diagram of the flagged Clifford circuit showing both the purification ($p_c^\text{ent}$) and absorbing ($p^{\rm abs}_c$) transitions, as a function of $p$ and of the inverse onsite Hilbert space dimension $(q+1)^{-1}$.}
    \label{fig:phase-diagram}
\end{figure}

We finally present our numerical results for the flagged Clifford model introduced in Sec.~\ref{sec:clifford_model}. The computations are performed using standard techniques, cf. e.g. Ref.~\cite{PhysRevB.100.134306, li2021statistical} for details.

We begin with the schematic phase diagram in Fig.~\ref{fig:phase-diagram}.
The two parameters of the model are the measurement rate $p$, and the onsite Hilbert space dimension $q+1$, respectively.
The absorbing transition occurs at ${p^{\rm abs}_c = \tilde p_c \simeq 0.3553}$ [cf. Eq.~\eqref{eq:pc}] for all values of ${q+1}$.
The entanglement transition coincides with the absorbing transition only at ${q+1 \to \infty}$, and is separated from the latter for any finite $q+1$.

The phase diagram is only schematic because ${q+1}$ is discrete. An additional subtlety is that the universality class of the MIPT can vary with $q+1$. In Clifford circuits, it is standard to take ${q+1 = q_0^M}$~\cite{gottesman1999higherdim}, where $q_0$ is a prime number and $M$ is an integer.
Thus, to have a series of different $q$, we could either (i) increase the value $q_0$ while fixing $M = 1$, or (ii) fix $q_0$ and increase $M$.
As discussed in Ref.~\cite{zabalo2022operator, li2021statistical}, the two series are not equivalent, as the symmetry group of the associated effective statistical mechanics model -- which determines the universality class of the MIPT -- depends on $q_0$, but not on $M$. This subtlety is specific to Clifford circuits: for Haar-random circuits, the MIPT universality class is believed to be independent of $q$.

From the point of view of the renormalization group, the simplest choice would be to study   the series (ii), so that all the  all examined points on the blue line in Fig.~\ref{fig:phase-diagram}  correspond to the same universality class. However, to access ${q+1 = q_0^{M > 1}}$, one  needs to group $M$ elementary qudits of dimension $q_0$ to form a large qudit of dimension $q$.
Numerically,
this would reduce the largest accessible system size by a factor of $M$, which is rather undesirable.

For this practical reason, we consider series (i), where we fix ${q+1 = q_0}$ and increase the prime $q_0$, which introduces only minimal numerical overhead.
Thus,  the points on the blue line in Fig.~\ref{fig:phase-diagram} that we examine, for different $q+1$, in fact correspond to distinct universality classes of MIPT, albeit with broadly similar scaling properties (e.g. conformal invariance). Despite this subtlety, our choice of values for $q+1$ can still illustrate our main point, namely the separation of the MIPT and absorbing transitions at finite values of $q+1$, 
and the crossover between the two transitions at large values of $q+1$.
Previously this method was used to probe such a crossover in MIPTs without feedback~\cite{li2021statistical}.

After this digression, we can now discuss numerical evidence for the phase diagram in Fig.~\ref{fig:phase-diagram}.

\subsection{Absorbing transition and the $q+1 \to \infty$ limit}
\label{sec:absorbingqinfinity}

\begin{figure}[t]
    \centering
    \includegraphics[width=.45\textwidth]{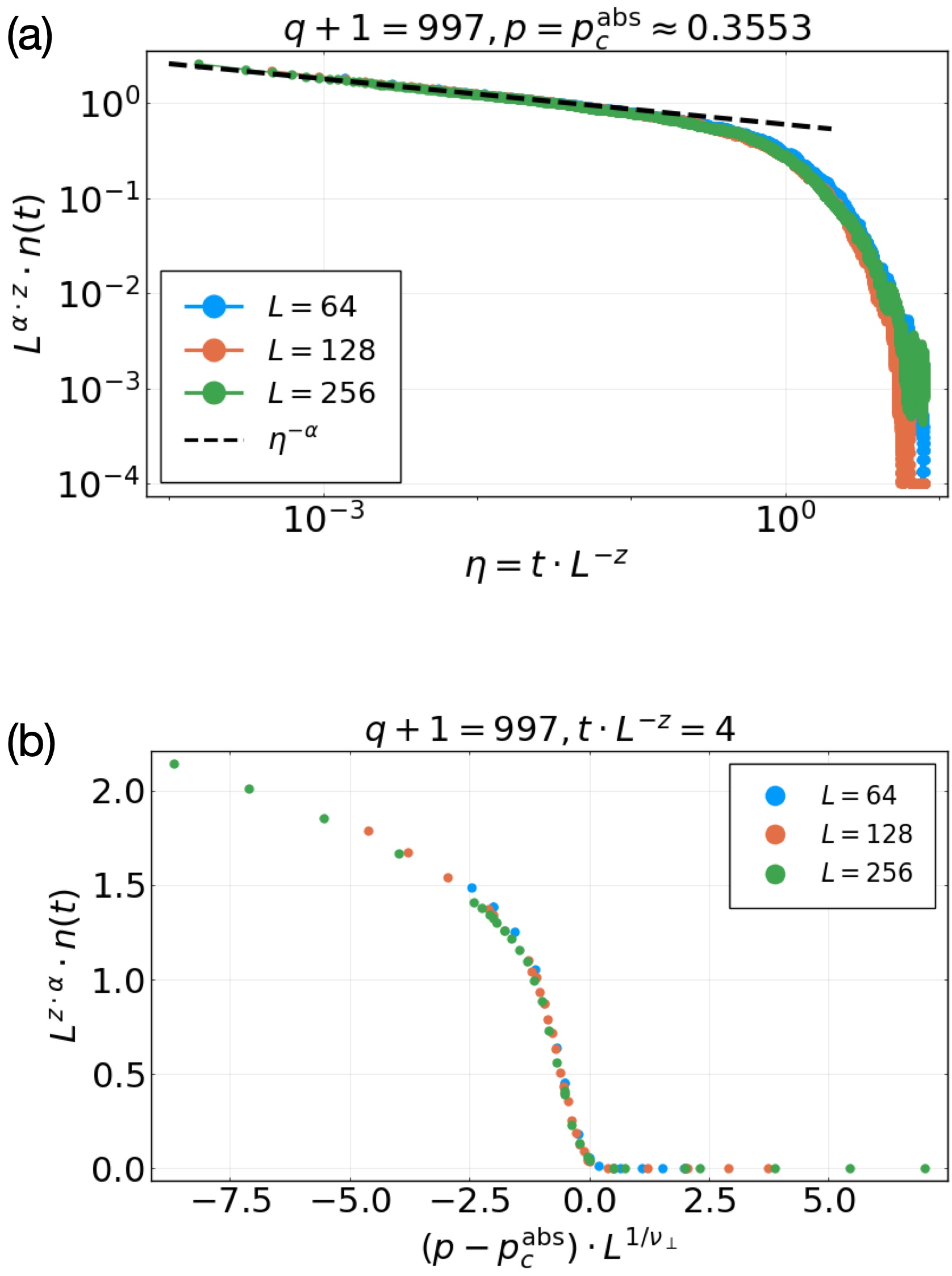}
    \caption{Numerical results for the order parameter of the absorbing transition, $n = \langle \mathcal{P} \rangle$, from a direct simulation of the flagged Clifford circuit at $q+1 = 997$.
    The two panels are different cross sections of the scaling function $\widetilde{F}$ in Eq.~\eqref{eq:P-critical-scaling-form}, at $(p-p^{\rm abs}_c)\cdot L^{1/\nu_\perp} = 0$ and at $\eta = t \cdot L^{-z} = 4.0$, respectively.
    }
    \label{fig:q-997-density}
\end{figure}

As  mentioned in the previous section, the order parameter for the absorbing transition, ${n=\langle\mathcal{P}\rangle}$, is proportional to the flag density $n^{\rm cl}$, giving the following standard scaling form near the critical point~\cite{henkel2008non}
\begin{equation}
\label{eq:density-scaling-form-DP}
    n(t,L,p)  
    = t^{-\alpha} \cdot F\left(\frac{t}{\xi_\parallel}, \frac{L}{\xi_\perp}\right).
\end{equation}
Here, $t$ is the time (i.e. circuit depth), $L$ is the length of the qudit chain, $F$ is a universal function, and $\alpha$ is a critical exponent of directed percolation.
The variables
$\xi_\parallel$ and $\xi_\perp$ are the previously introduced time and space correlation lengths, respectively.
As mentioned, they diverge near the critical point as
\begin{equation}
    \xi_{\parallel,\perp} \propto |p-p^{\rm abs}_c|^{-\nu_{\parallel,\perp}}, 
\end{equation}
where 
\begin{equation}
z = \nu_\parallel / \nu_\perp
\end{equation}
is the dynamic exponent of DP. We have $z>1$, signaling the anisotropy of the DP cluster.
For reference, the numerical values of the critical exponents are~\cite{henkel2008non}
\begin{equation}\label{eq:exponents}
    \alpha \approx 0.159, \ 
    \nu_\parallel \approx 1.733,\ 
    \nu_\perp \approx 1.097,\ 
    z \approx 1.581.
\end{equation}
We may rearrange the variables in the scaling form Eq.~\eqref{eq:density-scaling-form-DP} so that
\begin{equation}
\label{eq:P-critical-scaling-form}
    n(t) = L^{-z \cdot \alpha} \widetilde{F}\left( (p-p^{\rm abs}_c)\cdot L^{1/\nu_\perp}, \ \eta \coloneqq t\cdot L^{-z}\right),
\end{equation}
where $\eta$ sets the dimensionless aspect ratio and
\begin{align}
    \widetilde{F}(0, \eta) \propto \eta^{-\alpha}, \quad \text{ as } \eta \to 0.
\end{align}
We numerically verified Eq.~(\ref{eq:P-critical-scaling-form}), as shown in Fig.~\ref{fig:q-997-density}. We also report a typical snapshot of the classical-flag dynamics in Fig.~\ref{fig:dp-config}, both below, at, and above the critical rate $p^{\rm abs}_c$.

\begin{figure}[t]
    \centering
    \includegraphics[width=.44\textwidth]{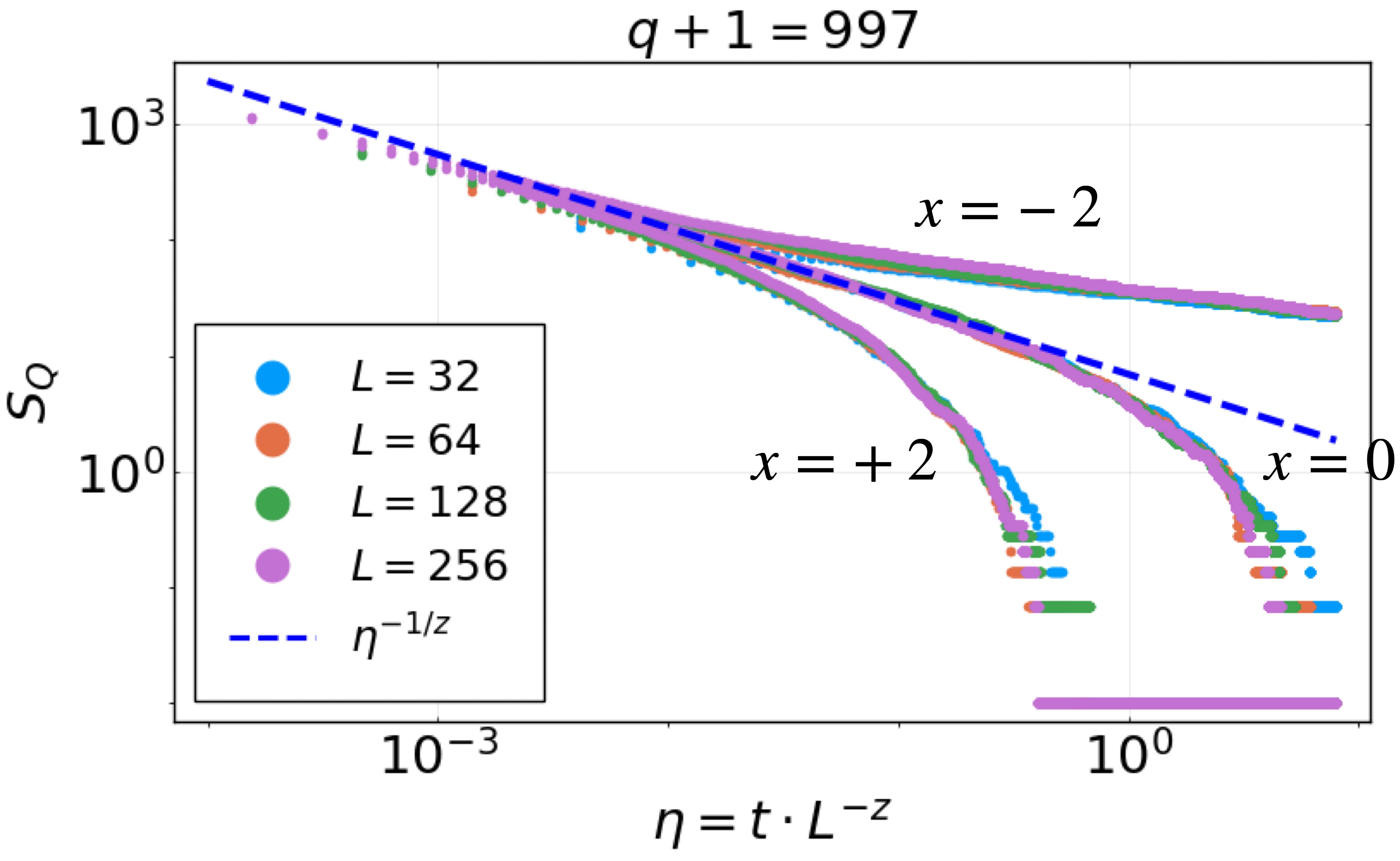}
    \caption{Scaling collapse of the entropy $S_Q$ starting from a maximally mixed initial state, for large $q+1=997$, using DP exponents. We plot different cuts of the scaling function~\eqref{eq:SQ_pc_A}, with fixed $x = (p-p^{\rm ent}_c)\cdot L^{1/\nu_\perp}$ and $p_c^{\rm ent} \approx \tilde p_c$ in Eq.~\eqref{eq:pc}. For $q+1 = 997$, the purification transition effectively coincides  with the DP transition, as  the separation between the two cannot be resolved numerically.}
    \label{fig:q-997-S-purification}
\end{figure}

As we have a good analytic understanding of the order parameter $n(t)$, from now on we focus on entanglement entropies averaged over trajectories, whose behavior is less obvious.
A nontrivial check of the phase diagram is the limit $q+1 \to \infty$
where the MIPT and the absorbing transition coincide, as the entanglement entropy saturates the minimal cut of the active DP tensor network~\cite{hayden2016holographic}.
In particular, with a maximally mixed initial state, we expect the entropy of the state (computed for each quantum trajectory, then averaged over trajectories, see Eq.~\eqref{eq:entropy}) to obey the following scaling form,
\begin{equation}
\label{eq:SQ_pc_A}
     S_Q(t,L,p)  = 
    \ln (q+1) \cdot
     G_{\rm DP}\left( (p-p^{\rm ent}_c)\cdot L^{1/\nu_\perp},\, t\cdot L^{-z}\right),
\end{equation}
where $p_c^{\rm ent}=p^{\rm abs}_c$ in this limit $q+1 \to \infty$, coinciding also with $\tilde p_c$ in~\eqref{eq:pc}.
Here the we use $S_Q$ to denote the mixed-state entropy $S$ from Eq.~\eqref{eq:entropy}.
The subscript $Q$ denotes the set of all $L$ qudits, rather than any subset of them.
We confirm this scaling form by numerical results at $q+1 = 997$, as shown in Fig.~\ref{fig:q-997-S-purification}.
At $q+1 = 997$ the distance between the two transitions, $\delta p = |p_c^{\rm ent} - p^{\rm abs}_c|$, is too small to be numerically resolvable.

\subsection{Entanglement transition at finite $q+1$}

\begin{figure}[t]
    \centering
    \includegraphics[width=.45\textwidth]{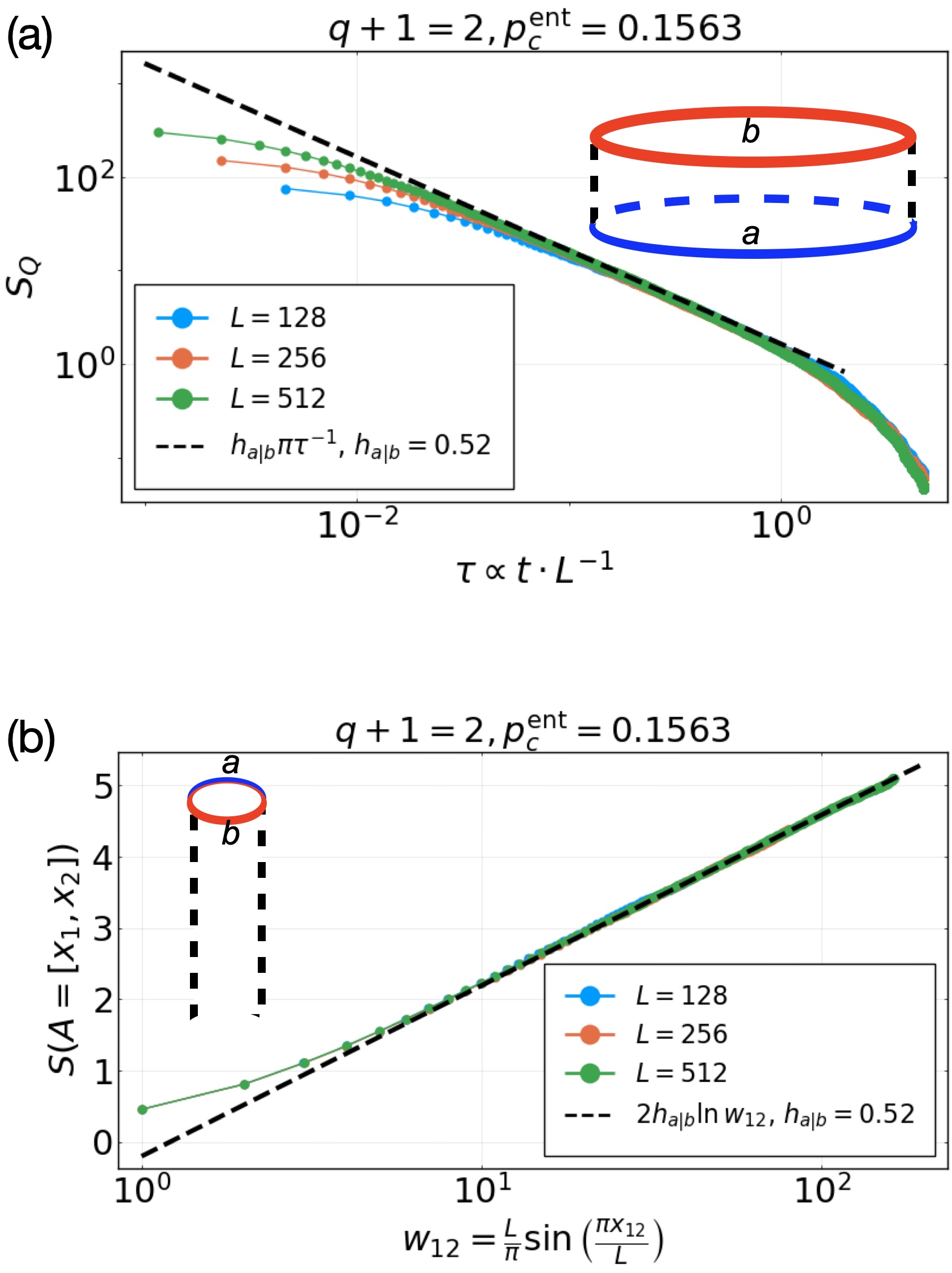}
    \caption{Entanglement scaling and conformal invariance at the MIPT (purification transition). (a) Purification of  a maximally mixed initial state with periodic spatial boundary condition in the regime $t \ll L$. The entropy $S_Q$ of the state obeys the scaling form~\eqref{eq:G_tau_small}, allowing us to extract the critical exponent $h_{a|b} \approx 0.52$.
    The value is consistent with previous findings, see Ref.~\cite{li2021conformal}.
    (b) Entanglement entropy of a subsystem $A = [x_1, x_2]$ in the steady-state $t \gg L$, obeying the scaling form~\eqref{eq:SA_critical_steady}.with the same exponent $h_{a|b}$. 
    The initial state is taken here to be a pure product state.}
    \label{fig:q-2-S-purification}
\end{figure}

We now turn to finite values of $q+1$. 
Recall 
that $p_c^\text{abs}$ in 
Fig.~\ref{fig:phase-diagram} is a $(q+1)$-independent constant that follows from the standard model of  DP, whereas the purification transition $p_c^{\rm ent}$ is pushed away from $p^{\rm abs}_c$ by $1/(q+1)$ fluctuations as we decrease $q$.
We mostly focus on ${q+1=2}$, 
where $|p^{\rm ent}_c - p^{\rm abs}_c|$ is largest, so that finite-size crossover effects are negligible.
We leave discussions of large values of $q$ and the associated finite-size crossover to Sec.~\ref{sec:crossover}.

A notable difference between the universal properties of the MIPT (at finite $q$) and of the absorbing transition is that the former are isotropic in spacetime, with dynamical exponent $z=1$, and also conformally invariant. Here we 
give evidence for this conformal invariance at $p=p_c^{\rm ent}$ for the value $q+1=2$. Conformal invariance also provides strong constraints on the functional forms of the entanglement entropies, and allows us to extract  critical exponents~\cite{li2021conformal}.
For the case of the purification of a maximally mixed initial state with periodic spatial boundary conditions, the entropy of the state follows the scaling form
\begin{equation}
\label{eq:SQ_pc_P}
 S_Q(t, L, p=p^{\rm ent}_c) = G^{(q+1)}_{\rm MIPT}(\tau),
\end{equation}
where $\tau = vt/L$ is the aspect ratio of the circuit, where $v$ is a model-dependent velocity that has to be determined separately,\footnote{\label{fn:conformal_mapping}This nonuniversal velocity is a property of the bulk of the circuit, whose value can be fixed independently of $h_{a|b}$ using a conformal mapping for a circuit with variable depths and open boundary conditions.
We refer the reader to Ref.~\cite{li2021conformal} for details.} and $G^{(q+1)}_\mathrm{MIPT} (\tau)$ is a universal scaling function with the following asymptotic form as $\tau \to 0$:
\begin{equation}
\label{eq:G_tau_small}
    G^{(q+1)}_{\rm MIPT}(\tau) = h_{a|b}\,  \pi \tau^{-1}, \quad \tau \ll 1.
\end{equation}
Here $h_{a|b}$ is a universal (boundary) critical exponent of the MIPT.
Both the scaling function 
 $G^{(q+1)}_{\rm MIPT}$ and the exponent $h_{a|b}$ can depend on $q+1$, for the reasons discusssed in Sec.~\ref{sec:Clifford_model_results}; we do not explicitly write this dependence for~$h_{a|b}$.
 
Moreover, when ${t/ L\rightarrow \infty}$, so that the system is in the steady state (which is pure), the entanglement entropy of a subsystem $A = [x_1, x_2]$ must have the following scaling form, as also dictated by conformal invariance,
\begin{equation}
\label{eq:SA_critical_steady}
    S_{A = [x_1, x_2]}(t \gg L, p=p^{\rm ent}_c) = 2h_{a|b} \ln \left(\frac{L}{\pi} \sin\left( \frac{\pi x_{12}}{L} \right)\right),
\end{equation}
where $h_{a|b}$ is the same exponent appearing in Eq.~\eqref{eq:G_tau_small}. 

For ${q+1=2}$, our numerical results are shown in Fig.~\ref{fig:q-2-S-purification}.
They are in good agreement with the scaling forms~\eqref{eq:SQ_pc_P}--\eqref{eq:SA_critical_steady}.
Here, we estimate $p^{\rm ent}_c \approx 0.1563$ from the best fit of numerical data to the scaling form in Eq.~\eqref{eq:SA_critical_steady}, 
shown in Fig.~\ref{fig:q-2-S-purification}(b),
and $v \approx 0.59$ using the method based on conformal mapping, described in footnote~\ref{fn:conformal_mapping} and Ref.~\cite{li2021conformal}.
Moreover, the data appears consistent with a  value of $h_{a|b} \approx 0.52$, 
close to that of the $q+1=2$ MIPT without feedback~\cite{li2021conformal},
 suggesting that the entanglement transitions in the adaptive and non-adaptive models are in the same universality class.

We have also verified conformal invariance at ${q+1 = 3, 5}$, and find that the results are again consistent with values of $h_{a|b}$ found previously~\cite{li2021statistical} (data not shown).

\subsection{Finite-size crossover behavior \label{sec:crossover}}

When $q$ is large but finite there is a crossover between a regime at smaller lengthscales, where entropies are set by the properties of directed percolation clusters, and the asymptotic large scale regime which shows more generic MIPT behavior. 

First,  consider the simpler setting where $q=p_0^M$, with a fixed prime $p_0$ (see the discussion at the beginning of Sec.~\ref{sec:Clifford_model_results}).
When $M$ is  large, ${\delta p = |p^{\rm ent}_c - p^{\rm abs}_c|}$ is small. The simplest conjecture is that  there is a single lengthscale controlling the crossover, together with the associated timescale.
Then  $S_Q$ is described by the following universal crossover scaling form at large~$t$,~$L$
\begin{equation}
\label{eq:crossover-scaling-function}
    S_Q(t, L, p=p^{\rm ent}_c) = \Phi\left(
    \eta_t \coloneqq \frac{t}{\xi_\parallel^\star},
    \eta_x \coloneqq  \frac{L}{\xi_\perp^\star}\right),
\end{equation}
where for  simplicity we have set $p$ equal to the location of the MIPT, and where 
${\xi_{\parallel,\perp}^\star = |p^{\rm ent}_c - p^{\rm abs}_c|^{-\nu_{\parallel, \perp}}}$ are the correlation time/length scales determined by the distance between this point and the absorbing transition.
These may be thought of as the crossover time/length scales:
assuming that ${p_c^\text{ent}-p_c^\text{abs}\sim 1/q}$ as in the Haar case (Sec.~\ref{sec:MIPT}), then ${\xi_{\parallel,\perp}^\star \sim q^{\nu_{\parallel, \perp}}}$.
We expect the following limiting behaviors of the scaling function,
\begin{align}
\label{eq:crossover_scaling_function_limits}
    \Phi (\eta_t, \eta_x) = \begin{cases}
    \log (q+1) \cdot 
        \widetilde{G}_{\rm DP} (\eta_t / \eta_x^z) & \text{for }\eta_t,\,\eta_x \to 0,\\
        \widetilde{G}_{\rm MIPT}(\eta_t / \eta_x) & \text{for } \eta_t,\,\eta_x \to \infty.
    \end{cases}
\end{align}
The functions $\widetilde{G}_{\rm DP}$ and $\widetilde{G}_{\rm MIPT}$ are those in Eqs.~\eqref{eq:SQ_pc_A}, \eqref{eq:SQ_pc_P}, respectively, up to 
$O(1)$ prefactors in the argument.
They represent the different critical behaviors one expects to observe when the size of the tensor network is below or above the crossover scales.
We note that the scaling functions $\Phi$ and $\widetilde G_{\rm MIPT}$ can depend on the prime~$p_0$.

\begin{figure}[t]
    \centering
    \includegraphics[width=.45\textwidth]{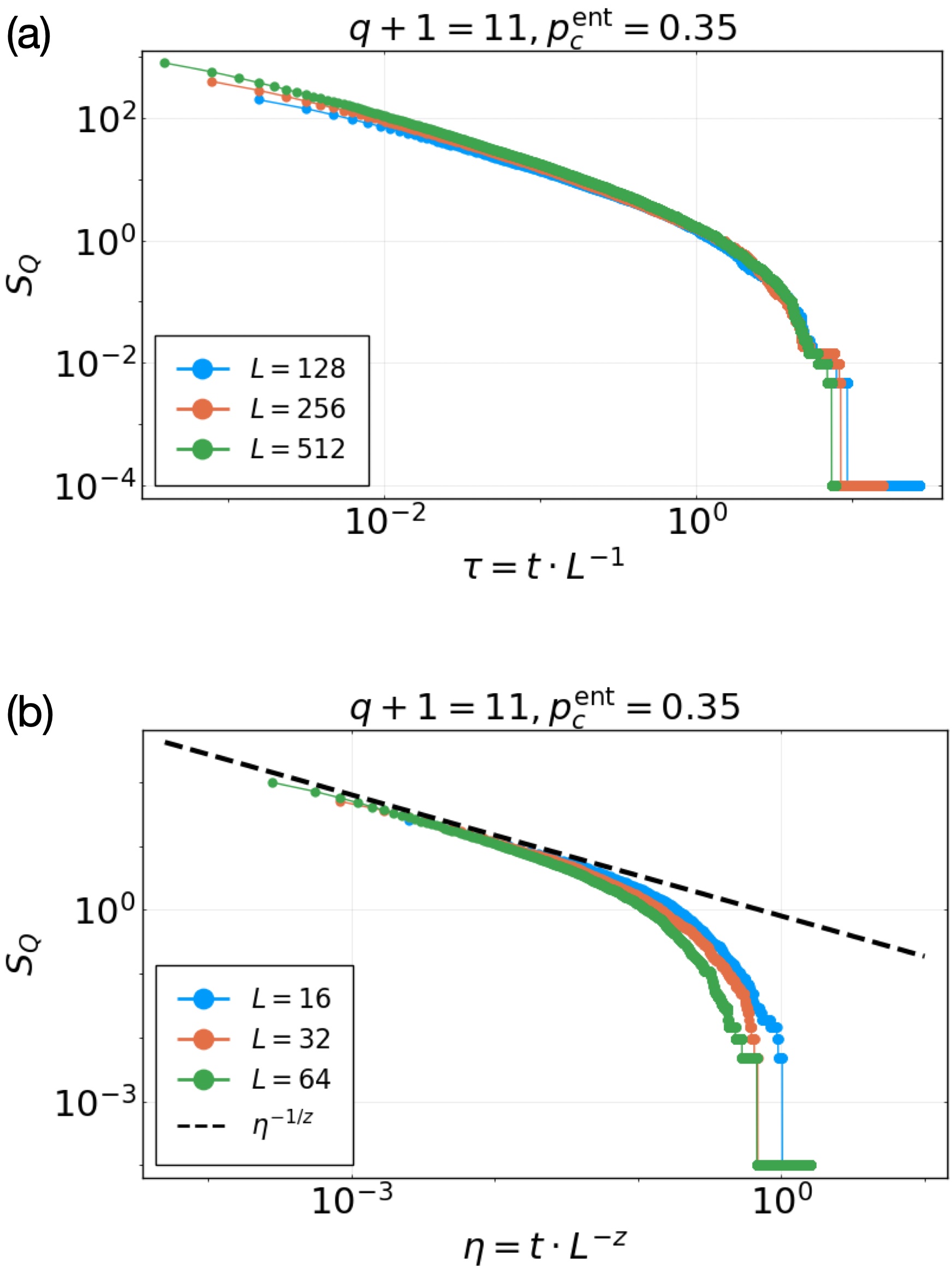}
    \caption{Numerical results for the crossover, for relatively large $q$, 
    between DP behavior at small scales and generic MIPT behavior at large scales (Sec.~\ref{sec:crossover}). 
    We focus on ${q+1=11}$, and tune to the critical point of the purification transition, $p = p_c^{\rm ent} \approx 0.350$.
    For larger system sizes $L$ (panel a), and at long times, we have $\eta_t \gg 1$ and $\eta_x \gg 1$ (with $\eta_t \coloneqq t / \xi_\parallel^\star$, $\eta_x \coloneqq x / \xi_\perp^\star$).
    In this regime we expect 
    $S_Q\simeq \widetilde{G}_{\rm MIPT}(\eta_t / \eta_x)$ as supported by numerics.
    For smaller system sizes $L$ (panel b), and at short times, we have $\eta_t \ll 1$ and $\eta_x \ll 1$.
    In this regime the entire system behaves like a critical cluster of directed percolation, with entanglement entropies determined by minimal cuts through the cluster.
    We expect
    $S_Q = \widetilde{G}_{\rm DP}(\eta_t / \eta_x^z)$, again consistent with numerical results. Note the distinct values for the dynamical exponent in the two regimes. 
    }
    \label{fig:S-purification-crossover}
\end{figure}

The case we are studying numerically here, where $q$ is a prime, is more subtle, 
since the infra-red universality class depends on $q$, but we expect a qualitatively similar crossover between a regime with scaling exponents (for example the dynamical exponent $z$) that are set by directed percolation, and a regime with a dynamical exponent of unity.
We numerically verify these limiting behaviors in Eq.~\eqref{eq:crossover_scaling_function_limits} at $q+1=11$ and plot the results in Fig.~\ref{fig:S-purification-crossover}.
We first locate the purification transition $p_c^{\rm ent}$ by looking at larger system sizes $L = 128, 256, 512$ (where  $\eta_x \gg 1$), where we observe that $S_Q$ collapses onto a function of the aspect ratio $\tau = t/L$ when $t$ is sufficiently large, as in the case ${q+1=2}$.
Some deviations are seen when $t$ is small, as expected.
For the smaller system sizes $L = 16, 32, 64$ (where $\eta_x \ll 1$) and for short times we observe that $S_Q$ instead depends on the anisotropic scaling variable  $t/L^z$, again as in Eq.~\eqref{eq:crossover_scaling_function_limits}.
The collapse breaks down at long times (when $t$ gets large), again as expected.

Above, we took the qudit dimension to be a fixed prime, ${q+1=q_0}$. It would also be interesting to set ${q+1=q_0^M}$ and to vary $M$, giving an additional tuning parameter. 
This would allow us to extract the exponent $\mu$ governing the crossover scale, ${\xi_\perp^\star(q)\sim q^\mu}$. It would also be interesting to see whether data for different $M$ could be collapsed using Eq.~\eqref{eq:crossover_scaling_function_limits}.

A striking difference between the two kinds of transition is in the dynamic exponent $z$.
The absorbing transition is anisotropic in spacetime, with ${z>1}$,
while 
the MIPT at finite $q$ is asymptotically conformally invariant
with ${z=1}$. The crossover scaling form Eq.~(\ref{eq:crossover_scaling_function_limits})  implies that the nonuniversal speed $v$ at the MIPT tends to zero as  ${q\to\infty}$ in this model, but it is finite for any finite $q$.

\section{Discussions \& generalizations}
\label{sec:conclusions}

In this paper, we have introduced a simple class of models exhibiting both a measurement-induced entanglement transition in quantum trajectories and an absorbing-state transition that is in the DP universality class. Measurements in these models define an ETN whose entanglement properties determine the purification (entanglement) transition. This allows us to show that the entanglement and absorbing-state transitions are generically distinct and unrelated to each other, except in the limit of infinite onsite Hilbert space dimension. By formulating a Clifford version of these models using flagged qudits, we were able to verify those predictions and to analyze the finite-time crossover between the two transitions numerically. 

Our theoretical analysis relied on the simplification that, in the models considered, the absorbing-state transition can be directly related to the connectivity of the ETN associated with a typical quantum trajectory. Heuristically, we now argue that this logic extends to more general models with a DP transition into an absorbing pure product state.
   
In each of the models defined in Sec.~\ref{sec:the_model}, a simple microscopic rule for ``flagging'' bonds was sufficient to define an ETN with two  properties: 
(1) the connectivity phase transition of the ETN coincided with the physical absorbing-state transition; and (2) the ETN faithfully reproduced the true quantum dynamics.
In the first model, the flags were set by directly measuring the occupancy of every bond.
In the second model, where the experimentalist had less measurement information,
the  flags at a given time  also took the previous time-step's flags into account.

For more general models these microscopic rules are not sufficient.
As an example, consider a model in which the measurement and resetting operations  take place only on even-numbered sites, $j\in 2\mathbb{Z}$. 
One can check that it is still possible to have an absorbing state transition in such a model.\footnote{\label{footnote:weak-measurement-model-updates}Let the probability of a given even-numbered site being reset in a given time step be $p$. 
Let the probability that an arbitrary site is occupied at a given time $t$, prior to the resetting operations, be $\langle\mathcal{P}\rangle_t$ (if we take  a brickwork pattern of Haar-random unitaries, this probability is the same for even and odd $j$). Then by bounding $\langle\mathcal{P}\rangle_{t+1}$ in terms of  $\langle\mathcal{P}\rangle_t$ and $p$, we can check that for sufficiently small ${1-p}$ the occupation probability decays exponentially with time, implying that the system is in the inactive phase. A more detailed discussion and simulation of this model is given in Appendix~\ref{sec:markov_weak}.} But if we continued to use the same protocol as in Sec.~\ref{sec:clifford_model} to define the ETN, then bonds with odd $j$ would always be flagged as active, so that the ETN would fail to show a connectivity transition.

To rectify this, we can imagine a more coarse-grained construction of the ETN in more general models. 
For concreteness, assume the model is such that each trajectory involves measurements on an order-one fraction of the spacetime bonds.

Recall that if a site is measured and found to be empty, we can eliminate the corresponding bond from the ETN. On the other hand, an \textit{unmeasured} site in general has a nonzero probability amplitude to be occupied, and so in general cannot be eliminated from the ETN without inducing some \textit{error} in the quantum state represented by the tensor network. 
However, we now argue that, close to the DP transition, the occupation probability is exponentially small for many of the unmeasured bonds. As a result these bonds can be eliminated from the ETN with only a very small error. Once this is done we can repeat (at least at a heurstic level) the argument from Sec.~\ref{sec:MIPT} showing that the two transitions are separated.

When we are  close to the DP transition, on the active side, 
there will exist large spacetime regions (with spatial sizes up to order $\xi_\perp$ and temporal sizes up to order $\xi_\parallel$) 
inside of which all of the \textit{measured} bonds are found to be empty.
Generically, an unmeasured bond has some nonzero amplitude to be active, and some nonzero amplitude to be inactive. However, deep in such a region, this amplitude will be exponentially small in the distance to the active region.
(We discuss this for the example of the model with measurements on even sites in App.~\ref{sec:markov_weak}.)

Therefore we expect that if in each trajectory we simply eliminate these bonds from the ETN, 
keeping  a collar of size $\sim R$ (with ${1\ll R \ll \xi_\perp}$) around the active regions,
then we will perturb the location of the entanglement transition only by an amount that is exponentially small at large $R$.
This then gives an ETN with the key property that its directed percolation transition coincides with the absorbing state transition, 
allowing us to reapply the argument of Sec.~\ref{sec:arguments} for the separation of the two transitions.
Making this argument precise would require slightly more care, in particular in estimating the cost of a ``red bond''.\footnote{This cost should no longer be estimated simply using the min cut formula for the ETN. The min-cut formula would give something of order $R$ (for $R\gg 1$), because of the collar region we have included. This is an overestimate, because most of the bonds in the collar region have an exponentially small amplitude to be occupied, and so contribute negligibly to the cost of the entanglement domain wall. The correct scaling is presumably of order~$R^0$.}

In our work we have considered the setting where the absorbing state has zero entropy, so that trajectories are  manifestly disentangled in the inactive phase.  The above analysis does not apply to more general absorbing states with non-trivial entanglement scaling.
Let us briefly comment on some models of this type. 

It is  possible  to construct models with transitions that are in the DP universality class, but where the averaged density matrix has extensive entropy on both sides of the transition.
A trivial way to do this is to take one of the adaptive qudit chain models discussed above and``stack'' it with a non-adaptive chain, 
giving a two-leg ladder geometry
 (weak interactions can be switched on between the two legs so long as they preserve the zero-occupation number state of the first leg).
Tuning both the  rate of measurement/control operations in the first leg and the rate of projective measurements in the second leg
gives a two dimensional phase diagram with both a DP transition line and an MIPT transition line.
In general the DP transition is unrelated to the MIPT transition,
and can occur within either the entangled or the disentangled phase.\footnote{By tuning two parameters we can access a point where the two kinds of transition cross. In this case the critical DP configurations can be thought of as a source of power-law correlated disorder for the entanglement degrees of freedom, which can change the universality class of the MIPT at this point.}

Finally, let us conclude with some outlook on future work. 
Adaptive dynamics of the type considered here (see also Refs.~\cite{buchhold2022revealing,  friedman2022measurement}), featuring local feedback, 
do not make it possible to  probe the entanglement MIPT using ``conventional'', non post-selected measurements.
The experimental observation  of the MIPT likely requires either heavy postselection~\cite{koh2022experimental}, or supplementary (nonlocal) classical computation and decoding using measurement outcomes~\cite{noel2022measurement} --- see Refs.~\cite{GullansNN,BarrattLearning,XEntropyYaodong,garratt2022measurements,feng2022measurementinduced} for recent efforts in that direction.
Indeed, quantum correlations can be generated nonlocally along postselected trajectories~\cite{nahum2020entanglement,li2021conformal,yoshida2021decoding,sang2022ultrafast},  whereas correlations of observables in the averaged mixed state strictly obey the Lieb-Robinson bound.

However, even with only local feedback, adaptive quantum dynamics 
may reveal other  universal phenomena that are 
interesting in their own right and also experimentally accessible~\cite{chertkov2022characterizing}.
Adaptive operations can also arise naturally as simplifications of the effect of an environment, mimicking for example the decay of an excitation (note that the quantum automaton circuits whose MIPTs have been studied in Refs.~\cite{iaconis-2020-automata, knolle-2022a-MIPTclassical, knolle-2022b-MIPTclassical} also fall into the category of adaptive dynamics).

So far in adaptive dynamics we have mostly found critical phenomena that
(if we only have access to the averaged density matrix, rather than to trajectories)
are essentially classical: 
that is,  at large timescales they typically admit a description in terms of a simple classical stochastic process for the slow degrees of freedom. (Diffusion-annihilation of anyons
gives counterexamples, in the form of processes with  slow degrees of freedom that are ``non-classical'' and involve long-distance entanglement~\cite{nahum2020entanglement,lin2021reaction}.)
Finding examples of nonclassical dynamical phase transitions in adaptive quantum  dynamics is an interesting challenge for the future.\\

\noindent\textit{Note Added.} While finalizing this manuscript, we became aware of a related work which recently appeared on the arXiv~\cite{KhemaniInteractive}, where the authors analyze  the interplay of absorbing-state and entanglement transitions in Haar-random circuit models (see also Ref.~\cite{2022arXiv221105162R} for earlier related results). We also became aware of another closely related Clifford model by Piotr Sierant and Xhek Turkeshi, which appeared in the same arXiv posting~\cite{sierant2022controlling}.\\

\noindent\textit{Acknowledgements.} We thank Sebastian Diehl, Xiao Chen, Zhi-Cheng Yang, Nick O'Dea, Alan Morningstar, and Vedika Khemani for helpful discussions. This work was supported by the US Department of Energy, Office of Science, Basic Energy Sciences, under Early Career Award No. DE-SC0019168 (R.V.), and the Alfred P. Sloan Foundation through a Sloan Research Fellowship (R.V.).
Y.L. is supported in part by the Gordon and Betty Moore Foundation’s EPiQS Initiative through Grant GBMF8686, and in part by the Stanford Q-FARM Bloch Postdoctoral Fellowship in Quantum Science and Engineering.
L.P., Y.L. and R.V. acknowledge the hospitality of the Kavli Institute for Theoretical Physics at the University of California, Santa Barbara (supported by NSF Grant PHY-1748958), during the program ``Quantum Many-Body Dynamics and Noisy Intermediate-Scale Quantum Systems''.

\appendix

\section{Markov process for restricted measurements}
\label{sec:markov_weak}

In this Appendix, we provide more details for the discussions sketched in Sec.~\ref{sec:conclusions}, where we argued that our conclusions are more general than the specific models introduced in Sec.~\ref{sec:the_model}. We detail in particular the model introduced therein, cf. footnote~\ref{footnote:weak-measurement-model-updates}, deriving the stochastic process for the particle densities and providing visualization for the resulting ETN. 
As mentioned, in this case the latter is not directly specified by the dynamics of the classical flags, since the measurements are only performed on a subset of qudits (in our case the even-numbered ones, see below).
In order to define the ETN, it is necessary also to elimininate bonds with a nonzero, but very small, occupation amplitude.

There is some freedom here in how to define a bond's occupation amplitude or occupation probability. For simplicity, we consider the expected onsite particle density at a given time on unmeasured (odd) sites, conditioned on the observed occupation numbers -- a binary, discrete-valued classical flag -- of each measured (even) site after each of the previous timesteps, but averaged over all possible trajectories that lead to the same observed configuration of binary classical flags (including the previous random unitary gates and also where the resetting operations were performed).
The resulting quantity can be easily computed within a simple stochastic process, when the previous binary classical flags are provided as input.
This expected local average density can then be viewed as a continuous-valued classical flag.
In principle, by imposing some cutoff ${\epsilon\ll 1}$ for the local density, we could use this flag as a criterion to eliminate bonds from the ETN (see the dark regions in Fig.~\ref{fig:dp-config-weak-meas} below). 
Here we simply discuss the effective stochastic process.  We believe the discussion below provides sufficient evidence for our basic point: typically, inside a large region where most measurements return ``unoccupied'', the occupation amplitudes of the unmeasured qudits are exponentially small.

Consider a one-dimensional array with $L$ qudits, each with Hilbert space dimension $(q+1)$.
The circuit evolution is similar to the model in Fig.~\ref{fig:sketch}b, with block-diagonal unitary gates arranged in a brickwork circuit, 
and dilute feedback operations, as well as single-site measurements of the operators $\{M_0 = \ket{0}\bra{0}, M_1 = \mathcal{P} = \openone - \ket{0}\bra{0}\}$ in each timestep.
The only difference is that  measurement and feedback operations are only performed on \textit{even-numbered} sites.
If we redefine the unit cells such that each contains two neighboring sites (an odd and an even site), then these measurements can be thought of as a ``partial''. For convenience, in each timestep we make the measurements of $\mathcal{P}$ prior to the resetting operations.

To describe the time evolution of the particle densities, it suffices to consider a classical Markov process with classical variables:
on each odd site we have a real number within $g_j \in [0,1]$, and on each even site we have a binary variable $f_j \in \{0,1\}$.
The variable $g_j$ represents the expectation value of $\mathcal{P}$ on qudit $j$, but the measurement is never physically taken since $j$ is odd;
and the variable $f_j$ represents the outcome of the measurement on qudit $j$.

\begin{figure*}
    \centering
    \includegraphics[width=\textwidth]{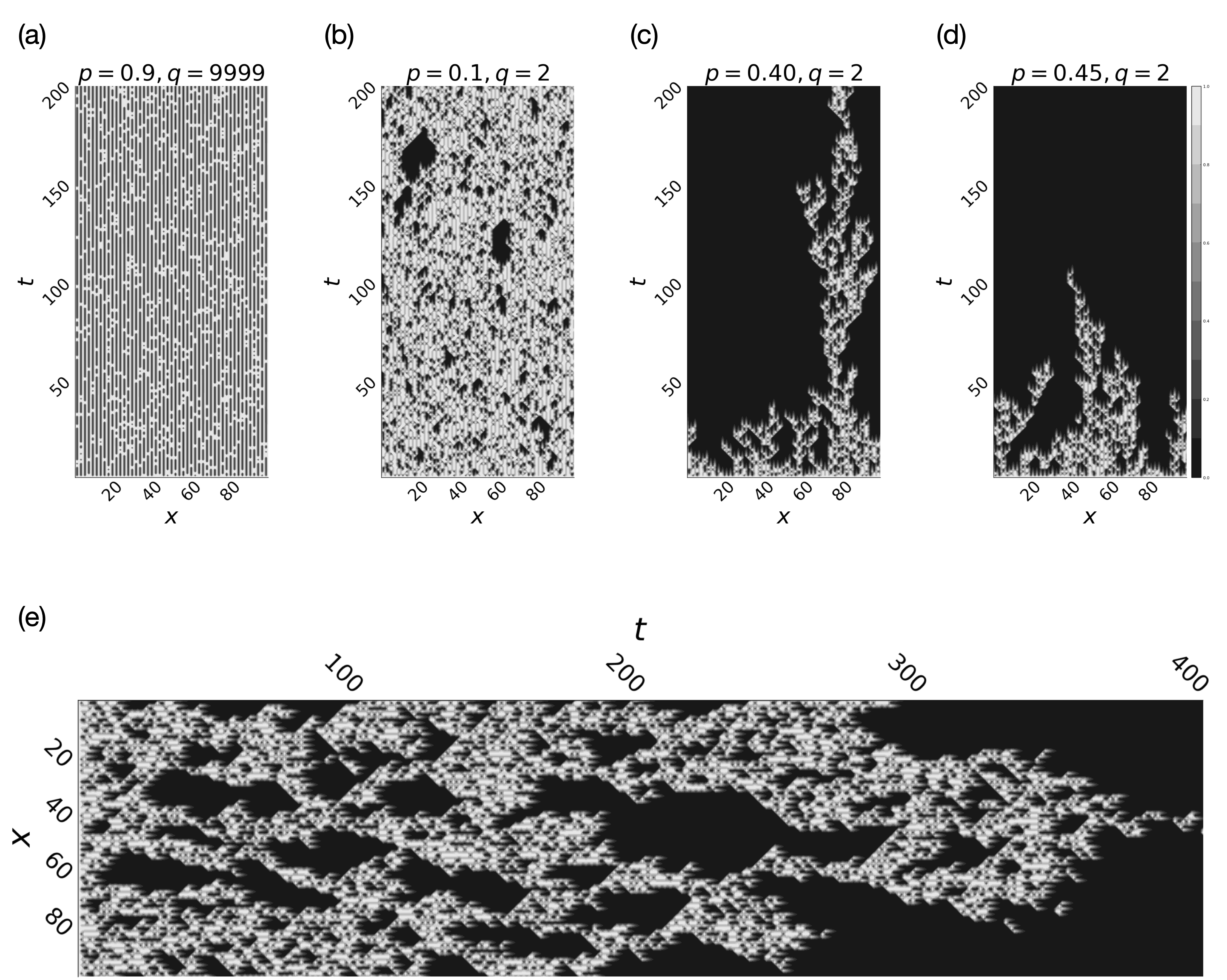}
    \caption{Configurations for the stochastic process defined in Eqs.~\eqref{eq:weak-meas-update-rule-combined-f-0},\eqref{eq:weak-meas-update-rule-combined-f-1}.
    The binary variable $f$ on even sites is in black $(f=0)$ or white $(f=1)$ color.
    The continuous variable $g \in [0,1]$ on odd sites are represented by gray levels.
    (a) As $q \to \infty$, the odd, unmeasured sites are going to be active $(g = 1)$ with probability $1$, even in the limit $p = 1.0$, and there is no connectivity transition (see Sec.~\ref{sec:conclusions}).
    Thus, to have a connectivity transition we have to be look at finite $q$.
    In (b-d), we plot the configurations at $q=2$ below, near, and above the transition.
    We see that there are sites that are ``gray''.
    There are regions in which all sites, whether directly measured or not, are inactive.
    (e) A configuration close to the critical point, on the active side.
    }
    \label{fig:dp-config-weak-meas}
\end{figure*}

The update rules for the pair $(f_j, g_{j+1})$ ($j$ even) can be derived as follows.
\begin{itemize}
    \item 
    $f_j = 0$.
    We write down the reduced state on qudits $j$ and $j+1$ as 
    \begin{align}
        \rho_{j,j+1} = \ket{0}\bra{0} \otimes \sigma_{j+1},
    \end{align}
    where
    \begin{subequations}
    \begin{align}
        \mathrm{tr} \left( \mathcal{P}_{j+1} \cdot \sigma_{j+1} \right) =&\ g_{j+1}, \\
        \mathrm{tr} \left( \ket{0} \bra{0} \cdot \sigma_{j+1} \right) =&\ 1-g_{j+1}.
    \end{align}
    \end{subequations}
    
    We consider the average output state when a random unitary as in Eq.~\eqref{eq:block_unitary} is applied, which should be diagonal
    \begin{align}
        \overline{\rho^U_{j,j+1}} 
        &= \ \mathbb{E}_U U \rho_{j,j+1} U^\dagger 
        = \ (1-g_{j+1})\ket{00}\bra{00} \nonumber\\ 
        &+ \frac{g_{j+1}}{q(q+2)} (\openone - \ket{00}\bra{00}).
    \end{align}
    
Following the random unitary gate, a measurement of $\mathcal{P}$ is performed on qudit $j$.
    We can calculate the probabilities for the two outcomes separately (recall that the projectors for the measurement are $\{M_0 = \ket{0}\bra{0}, M_1 = \mathcal{P} = \openone - \ket{0}\bra{0}\}$)
    \begin{subequations}
    \begin{align}
        p^{M_0} =&\ \mathrm{tr} [ \overline{\rho^U_{j,j+1}} \cdot (M_0)_j ] \nonumber\\
        &= (1-g_{j+1}) + \frac{q \cdot g_{j+1}}{q(q+2)}, \\
        p^{M_1} =&\ \mathrm{tr} [ \overline{\rho^U_{j,j+1}} \cdot (M_1)_j ] \nonumber\\
        &= \frac{q (q+1) \cdot g_{j+1}}{q(q+2)},
    \end{align}
    \end{subequations}
    and we have to update $g_{j+1}$ accordingly after the state is collapsed onto the measurement outcome,
    \begin{widetext}
    \begin{subequations}
    \begin{align}
        g_{j+1}^{M_0} =&\  \mathrm{tr} \left[ \mathcal{P}_{j+1} \cdot \frac{\overline{\rho^U_{j,j+1}} \cdot (M_0)_j}{\mathrm{tr} [ \overline{\rho^U_{j,j+1}} \cdot (M_0)_j ]} \right]
        =\ \frac{\frac{q \cdot g_{j+1}}{q(q+2)}}{ (1-g_{j+1}) + \frac{q \cdot g_{j+1}}{q(q+2)} } = \frac{g_{j+1}}{1+(1-g_{j+1})(1+q)}, \\
        g_{j+1}^{M_1} =&\  \mathrm{tr} \left[ \mathcal{P}_{j+1} \cdot \frac{\overline{\rho^U_{j,j+1}} \cdot (M_1)_j}{\mathrm{tr} [ \overline{\rho^U_{j,j+1}} \cdot (M_1)_j ]} \right]
        =\ \frac{\frac{q^2 \cdot g_{j+1}}{q(q+2)}}{ \frac{q(q+1) \cdot g_{j+1}}{q(q+2)} } = \frac{q}{q+1}.
    \end{align} 
    \end{subequations}

    Recall that, after measuring $M_1$ on qudit $j$, we perform with probability $p$ a resetting operation to restore the qudit to the state $\ket{0}\bra{0}$. Summarizing, the rules are as follows,
    
    \begin{subequations}
    \begin{align}
        \text{with prob. } p^{M_0}:&\ (0, g_{j+1}) \to (0, g_{j+1}^{M_0}) \\
        \text{with prob. } p \cdot p^{M_1}:&\ (0, g_{j+1}) \to (0, g_{j+1}^{M_1}) \\
        \text{with prob. } (1-p) \cdot p^{M_1}:&\ (0, g_{j+1}) \to (1, g_{j+1}^{M_1})
    \end{align}
    \end{subequations}
    
    To reduce the number of samples needed in the numerical simulation, we may ``combine'' the first two branches, i.e. average over the two possible ways for  $f_j$ to get to zero at the end of the timestep:
    \begin{subequations}
    \label{eq:weak-meas-update-rule-combined-f-0}
    \begin{align}
        \text{with prob. } p^{M_0}+ p \cdot p^{M_1} &:\
        (0, g_{j+1}) \to \left( 0, \frac{p^{M_0} \cdot g_{j+1}^{M_0} + p \cdot p^{M_1} \cdot g_{j+1}^{M_1} }{ p^{M_0}+ p \cdot p^{M_1} } \right) \\
        \text{with prob. } (1-p) \cdot p^{M_1} &:\
        (0, g_{j+1}) \to (1, g_{j+1}^{M_1})
    \end{align}
    \end{subequations}
    If we average the new value of $g_j$ over the two branches, weighted by their corresponding probabilities, we have 
    $\overline{g_{j+1}(t+1)} = \frac{q+1}{q+2} g_{j+1}(t)$, that is, it is monotonically decreasing when the neighboring classical flag $f_j = 0$.
    This is the key point.
    In an ``inactive'' subregion where the binary classical flags are zero, the value of $\overline{g_{j+1}}$ is typically exponentially small in the distance to the boundary of the inactive region. In this model the scaling is simple: if the two neighbors of site $j+1$ are unoccupied for a time $\tau$, then after this time $\overline{g_{j+1}}\leq \left(\frac{q+1}{q+2}\right)^{\tau}$.
    
    \item 
    $f_j = 1$.
    In this case we have
    \begin{align}
        \overline{\rho^U_{j,j+1}} 
        = \frac{1}{q(q+2)} (\openone - \ket{00}\bra{00}),
    \end{align}
    and
    \begin{subequations}
    \begin{align}
        p^{M_0} =&\ \mathrm{tr} [ \overline{\rho^U_{j,j+1}} \cdot (M_0)_j ] = \frac{q}{q(q+2)}, \\
        p^{M_1} =&\ \mathrm{tr} [ \overline{\rho^U_{j,j+1}} \cdot (M_1)_j ] = \frac{q (q+1)}{q(q+2)},
    \end{align}
    \end{subequations}
    
    \begin{subequations}
    \begin{align}
        g_{j+1}^{M_0} =&\  \mathrm{tr} \left[ \mathcal{P}_{j+1} \cdot \frac{\overline{\rho^U_{j,j+1}} \cdot (M_0)_j}{\mathrm{tr} [ \overline{\rho^U_{j,j+1}} \cdot (M_0)_j ]} \right] = 1, \\
        g_{j+1}^{M_1} =&\  \mathrm{tr} \left[ \mathcal{P}_{j+1} \cdot \frac{\overline{\rho^U_{j,j+1}} \cdot (M_1)_j}{\mathrm{tr} [ \overline{\rho^U_{j,j+1}} \cdot (M_1)_j ]} \right] = \frac{q}{q+1}.
    \end{align} 
    \end{subequations}
    Recall that, after measuring $M_1$ on qudit $j$ we perform with probability $p$ a ``feedback'' to restore the qudit to the state $\ket{0}\bra{0}$. Summarizing, the rules are as follows,
    \begin{subequations}
    \begin{align}
        \text{with prob. } p^{M_0}:\ (1, g_j) \to (0, g_{j+1}^{M_0}) \\
        \text{with prob. } p \cdot p^{M_1}:\ (1, g_j) \to (0, g_{j+1}^{M_1}) \\
        \text{with prob. } (1-p) \cdot p^{M_1}:\ (1, g_j) \to (1, g_{j+1}^{M_1})
    \end{align}
    \end{subequations}
    We may combine the first two processes as in the previous $f_j=0$ case, in a similar manner
    \begin{subequations}
    \label{eq:weak-meas-update-rule-combined-f-1}
    \begin{align}
        \text{with prob. } p^{M_0}+ p \cdot p^{M_1} &:\
        (1, g_{j+1}) \to \left( 0, \frac{p^{M_0} \cdot g_{j+1}^{M_0} + p \cdot p^{M_1} \cdot g_{j+1}^{M_1} }{ p^{M_0}+ p \cdot p^{M_1} } \right) \\
        \text{with prob. } (1-p) \cdot p^{M_1} &:\
        (1, g_{j+1}) \to (1, g_{j+1}^{M_1})
    \end{align}
    \end{subequations}
    \end{widetext}
    On average $\overline{g_{j+1}(t+1)} = \frac{q+1}{q+2}$, regardless of the value of $g_{j+1}(t)$.
    Thus, when the neighboring classical flag $f_j = 1$, it acts like a bath that puts $g$ in a local equilibrium.

\end{itemize}

We simulate this stochastic process, starting with the initial state $f_j = 1$ for all even $j$ and $g_j = 1$ for all odd $j$.
We plot several configurations in Fig.~\ref{fig:dp-config-weak-meas}.

\bibliography{refs}

\end{document}